\documentclass[%
  reprint,
 superscriptaddress,
 amsmath,amssymb,
 aps,
 prb,
]{revtex4-2}

\usepackage{graphicx}
\usepackage{dcolumn}
\usepackage{bm}
\usepackage{subfigure}

\graphicspath{ {./figs/} }

\DeclareUnicodeCharacter{0301}{}  
\begin{document}


\title{Diversity of behavior after collisions of Sn and Si nanoparticles found using a new Density-Functional Tight-Binding model}

\author{Andr\'es Ruderman}
\affiliation{
 Universidad Nacional de Córdoba, Facultad de Matemática, Astronomı́a, Fı́sica y Computación, Córdoba (X5000HUA), Argentina
}%
\affiliation{
 Consejo Nacional de Investigaciones Cientı́ficas y Técnicas (CONICET), Instituto de Fı́sica Enrique Gaviola, Córdoba (X5000HUA), Argentina
}%

\author{M. B. Oviedo}%
\author{S. A. Paz}%
\email{apaz@unc.edu.ar}
\author{E. P. M. Leiva}%
\affiliation{%
 Universidad Nacional de Córdoba, Facultad de Ciencias Quı́micas,
Departamento de Quı́mica Teórica y Computacional, Córdoba (X5000HUA),
Argentina
}%
\affiliation{%
 Consejo Nacional de Investigaciones Cientı́ficas y Técnicas (CONICET),
Instituto de Fisicoquı́mica de Córdoba (INFIQC), Córdoba (X5000HUA), Argentina
}%

\date{\today}

\begin{abstract}
We present a new approach to studying nanoparticle collisions using Density Functional based Tight Binding (DFTB). A novel DFTB parameterisation has been developed to study the collision process of Sn and Si nanoparticles (NPs) using Molecular Dynamics (MD). While bulk structures were used as training sets, we show that our model is able to accurately reproduce the cohesive energy of the nanoparticles using Density Functional Theory (DFT) as a reference. A surprising variety of phenomena are revealed for the Si/Sn nanoparticle collisions, depending on the size and velocity of the collision:  from core-shell structure formation to bounce-off phenomena.
\end{abstract}

\maketitle


\section{Introduction}

The emergence of lithium batteries in 1991 sparked a wave of innovation and research across various fields due to their versatile applications, ranging from mobile devices to electrical grids. Graphite has been the most commonly utilized anode material, known for its favorable cyclability and electronic conductivity but limited gravimetric capacity (372 mAhg$^{-1}$). To overcome graphite limitations, composites of tin and silicon, among others, have been proposed \cite{li2018carbon, xu2018facile}. In our laboratory, we conducted a study on a tin-silicon-graphite material (Sn-Si-$\mu$G) synthesized through a ball milling process. Our investigation revealed that these composite electrodes exhibit superior electrochemical performance compared to the individual components \cite{smrekar2020mapping}. We determined that neither tin nor graphite significantly contribute to charge storage in this composite \cite{ruderman2021unveiling}. Instead, tin plays a crucial role in stabilizing silicon nanoparticles, forming bimetallic core-shell nanoparticles with a silicon core and a tin shell.

Bimetallic nanoparticles (BNs) have garnered significant attention for their wide range of technological applications in fields such as catalysis, therapeutics, drug delivery, biosensors, energy storage, and more \cite{elemike_2019, lomeli-marroquin_2019, sivamaruthi_2019, minal_2020, arora_2020, padilla-cruz_2021, arif_2023, makada_2023, amiripour_2018, ansari_2018, sohrabi_2022, basavegowda_2020, mustielesmarin_2021, bruno_2023, tao_2023, qi_2022, lee_2020, bhiradi_2022, ramesh_2022, sabeeh_2021, sharma_2019, idris_2023}. Fine-tuning the composition and distribution of chemical elements grants BNs exceptional properties compared to nanoparticles comprising individual components. An in-depth understanding of BNs is essential to optimize existing technological applications and explore new ones, like lithium-ion batteries. Computational methods have played a crucial role in advancing this knowledge, providing detailed atomistic insights into BNs' processes and properties \cite{coviello_2022, farkas_2021, ferrando_2018, walle_2019, ferrando_2008a, ferrando_2016, delarosaabad_2021, ferrando_2015, palomares-baez_2017, nelli_2019, zhang_2020b, palomares-baez_2017}. Our previous work, as well as studies by others, have investigated the collision and coalescence of metallic nanoparticles as a simple mechanical method for synthesizing BNs \cite{paz_2011b, paz_2014, farigliano_2020, farigliano_2017, goudeli_2019, grammatikopoulos_2019}. Such studies provide valuable insights into the formation of BNs in various scenarios, including the ball milling synthesis mentioned earlier.

The coalescence and collision of nanoparticles represent highly complex fundamental mechanisms underlying nanoparticle growth. Simulating these processes poses significant challenges, as achieving a sufficiently large time scale to observe the final stages of coalescence while collecting statistical data on particle velocity and orientation requires the simulation of multiple collision events. Moreover, the evolution of the process involves various trapped states, making it challenging or even infeasible to simulate large systems at normal temperatures using conventional simulation algorithms \cite{paz_2014, paz_2015a, paz_2011b, farigliano_2020, farigliano_2017}. Consequently, resource-intensive \textit{ab-initio} models are often impractical, leading to the adoption of semi-empirical approximations such as EAM-type potentials as the classical approach \cite{goudeli_2019, grammatikopoulos_2019}. However, in this study, we present a novel approach to investigate nanoparticle coalescence using Density Functional based Tight Binding (DFTB). Our results demonstrate that DFTB offers an excellent balance between computational cost and precision, thereby enhancing predictive capabilities compared to conventional methods.

In this paper, we present the results obtained from molecular dynamics (MD) simulations exploring the collision process between tin and silicon nanoparticles. Our aim is to elucidate the morphology and characteristics of the resulting BNs under different collision conditions, mimicking the collisions that occur in a planetary ball mill, as observed in our aforementioned study. To study this system, we developed a new DFTB parameterization, employing bulk structures as training sets. Our model accurately reproduces the cohesive energy of nanoparticles, as determined by DFT, for both pure tin and silicon and the bimetallic case. Surprisingly, our simulations reveal a diverse range of phenomena arising from Si/Sn nanoparticle collisions, which depend on the collision's size and velocity. These phenomena include embedding, fragmentation, bouncing apart, and gentle contact between Si and Sn nanoparticles. We attribute the discovery of this diverse behavior to the utilization of the DFTB model.

\section{Computational methods}

\subsection{Density functional tight-binding}

The self-consistent density functional tight-binding (SCC-DFTB) method applied in this work has been extensively described in the literature, for a complete formulation we address to references~\cite{elstner_1998, frauenheim_2000, seifert_2007,
gaus_2011}.  It is based on the second-order expansion of the Kohn-Sham (KS) energy functional around a reference density, $n^0$, composed of neutral atomic species:
\begin{equation}\label{e_dftb}
    E_{\mathrm{DFTB}}=\sum_i^{\mathrm{occ}}\langle \Psi_i|\hat{H}^0|\Psi_i\rangle+\frac{1}{2}\sum_{\mathrm{AB}}^M \gamma_{\mathrm{AB}}\Delta q_{\mathrm{A}}\Delta q_{\mathrm{B}} + E_{\mathrm{rep}}
\end{equation}
The first term of the above equation corresponds to a Kohn-Sham effective Hamiltonian, $\hat{H}^0$, evaluated at the reference density. These matrix elements can be parameterized as only one-body or two-body terms and are described according to equation \ref{h0}.
\begin{align}
 \label{h0}
 H^0_{\nu\mu}=\begin{cases}
  \epsilon_\mu & \mathrm{if}\; \nu=\mu\\
  \langle \phi_{\nu}| -\frac{1}{2}\nabla^2+v_{\mathrm{eff}}\left[n_A^0+n_B^0\right]|\phi_{\nu}\rangle&\mathrm{if}\;\mu\in A,\; \nu\in B\;\mathrm{and} \;A\ne B\\
  0& \mathrm{otherwise}
 \end{cases}
\end{align}
$\phi_i$ forms a minimal Slater-type atomic basis centered on the atomic sites, where $S_{\nu\mu}=\langle\phi_{\nu}|\phi_{\mu}\rangle$. $n_A^0$ is the reference density of the neutral atom $A$, and $v_{\mathrm{eff}}$ is the effective Kohn-Sham potential, constructed from the superposition of neutral atom-centered densities. The pseudoatomic basis and density are obtained by solving the modified KS equation
\begin{equation}
    \left[-\frac{1}{2}\nabla^2+v_{\mathrm{eff}}+v_{\mathrm{conf}}\right]\phi_{\mu}=\epsilon_{\mu}\phi_{\mu}
\end{equation}
where $v_{\mathrm{conf}}$ is a confining which has the following form:
\begin{equation}
 \label{vconf}
 V_{\mathrm{conf}}(r)=\left(\frac{r}{r_0}\right)^{\sigma}
\end{equation}
$r_0$ and $\sigma$ are real numbers, which can be chosen differently for atomic
orbitals $\phi$ and densities $n^0$\cite{vandenbossche_2019a, vandenbossche_2019}.

The second term in equation~\ref{e_dftb} is the energy due to charge fluctuations, where \linebreak $\gamma_{AB}\left(U_A,U_B,|\mathbf{R}_A-\mathbf{R}_B|\right)$ is an analytical function that interpolates smoothly between onsite interactions $U_A=\gamma_{AA}$ or Hubbard parameter ($U$) and the bare Coulomb interaction at large separation. The values used for $U$ are computed by assuming that they are equal to those of the isolated atoms and have been calculated as the difference of the electron affinity and ionization energy for different orbital angular momenta. The quantity $\Delta q_A=q_A-q_A^0$ is the self-consistent induced Mulliken charge on atom $A$; for details see reference \cite{elstner_1998}.

The third term in equation \ref{e_dftb} corresponds to the repulsive term named $E_{\mathrm{rep}}$, and contains all the cumbersome terms related to the distance-dependent diatomic repulsive potential and the core electron effects, ion–ion repulsion terms, as well as some exchange–correlation effects. The repulsive energy is a sum of contributions of repulsive potentials
$V_{\mathrm{rep}}(r)$ from each atom pair: 
\begin{equation}
 \label{e_rep}
 E_{\mathrm{rep}}=\sum_{I<J}V_{\mathrm{rep}}(R_{IJ})
\end{equation}
where $I$ and $J$ run over the atom indices in the system, and $R_{IJ}$ is the
distance between pair of atoms. $V_{\mathrm{rep}}$ is generally considered to
be an empirical function that is determined by fitting data from a higher level
electronic structure calculations such as DFT. In this work, $V_{\mathrm{rep}}(R_{IJ})$ is represented by an exponential decay function and a polynomial function, given as
\begin{equation}
 \label{v_rep}
 V_{\mathrm{rep}}(R)=\begin{cases}e^{-a_1R+a_2}+a_3 & 0\le R<R_{\mathrm{min}}\\
 \displaystyle\sum_{i=2}^m c_i\left(R_{\mathrm{cut}}-R\right)^i & R_{\mathrm{min}}\le R < R_{\mathrm{cut}}\\
 0 & R_{\mathrm{cut}} \le R 
 \end{cases}
\end{equation}
with $m$ being the order of the polynomial, $R$ represents the interatomic distance and was chosen to be $8$. The $a_i$ parameters were fitted using Levenberg-Marquardt algorithm to reproduce the DFT energies for each crystal system and the $c_i$ coefficients were optimized through a least-square fit. Each parameter, $c_i$ and $a_i$ are tabulated in the Slater-Koster files as well as $R_{\mathrm{cut}}$.

All DFT calculations were performed within an atomic simulation environment (ASE)~\cite{larsen_2017} with GPAW software \cite{enkovaara_2010, mortensen_2005}. The GPAW package is a real space grid algorithm based on the projector-augmented wavefunction method \cite{blochl_1994} that uses the frozen core approximation. Coordinates of Si, SiSn and Si were downloaded from The Materials Project \cite{jain_2013} (mp codes 149, 117, 1009813). DFT calculations were performed using PBE (Perdew-Burke-Ernzerhof) exchange-correlation functional with a kinetic energy cutoff of 600 eV and the Brillouin zone integration was done with a $(8 \times 8 \times 8)$ Monkhorst–Pack k-point sampling grid.

\subsection{Molecular dynamics simulations}

Four different Si clusters with a number of 5, 10, 11 and 40 atoms were collided with a Sn particle of 110 atoms. The initial configurations for Si$_5$, Si$_{10}$, and Si$_{11}$ were obtained from the work of Fernandes {\it et al.} \cite{fernandes2018quantitative}, while the configuration for Si$_{40}$ from Wang \cite{wang2005optimally} {\it et al.}. The Sn$_{110}$ nanoparticle was obtained from a spherical cut of the bulk $\beta$ phase taken from the Materials Project database \cite{Jain2013}, as this phase is the most stable at 300 K. 
Simulations were conducted using a modified version of the GEMS code~\cite{paz_2020b}, which was interfaced with the DFTB+ code~\cite{hourahine_2020}.

\begin{figure}
\center{\includegraphics[width=\linewidth]{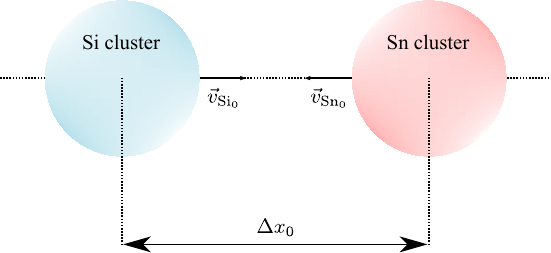}}
\caption{Model system for the collision between Sn and Si clusters. The distance between them, $\Delta x_0$ is three times bigger than the cutoff radio.
}
\label{part}
\end{figure}

Figure \ref{part} shows the model assumed for the collisions. Initially, the centers of mass of the clusters were aligned along the $x$ axis at a distance $\Delta x_0=30$\AA, which corresponds to a minimum distance between particles of 20\AA. This minimum distance ensures that the two colliding particles do not interact at the beginning of the simulation. The initial velocities were chosen so that the total momentum cancels out, i.e. $M_{\mathrm{Sn}}\vec{v}_{\mathrm{Sn}_0}=-M_{\mathrm{Si}}\vec{v}_{\mathrm{Si}_0}$, where $M_{\mathrm{Sn}}$ and $M_{\mathrm{Si}}$ are the masses of the Sn and Si clusters, respectively. In this way, the momentum of the center of mass (CM) is zero and, therefore, its velocity $\vec{v}_{\mathrm{cm}}=0$. As no external force acts over the system, $\vec{v}_{\mathrm{cm}}$ remains 0 and the position of the CM stands still. 

According to Chattopadhyay {\it et al.}, the available energy of a ball of mass $m$ in a ball milling is a function of the rotation frequency $\omega$ \cite{chattopadhyay2001mathematical}. At $\omega$=400 rpm, as used in our previous work \cite{smrekar2020mapping, ruderman2021unveiling}, the available energy per ball at the moment of impact was calculated to be $\sim$0.2 J \cite{chattopadhyay2001mathematical}. A 3 mm diameter stainless steel ball, as it was used in our work, has a mass of $1.1\times10^{-4}$ Kg which leads to a velocity at the moment of the impact of around 60 m/s. However, for the sake of running the present simulations in a reachable machine time, $v_{\mathrm{Sn}_0}$ and $v_{\mathrm{Si}_0}$ were chosen in such a way that: $v_{\mathrm{Si}_0}-v_{\mathrm{Sn}_0}$=1300 m/s, 1400 m/s and 1500 m/s. The study of collisions at lower velocities requires an accelerated simulation scheme and will be the subject of further works. Finally, for the three smallest Si clusters and each set of velocities ten different simulation runs were performed by rotating the clusters over the $x$, $y$, $z$ axis by a random angle. For the biggest Si particle, only 5 random rotations were studied.

The solvent accessible surface area (SASA) of the nanoparticles was computed using VMD \cite{HUMP96, VARSH1994}.  The solvent-accessible surface area (SASA) of the nanoparticles was calculated using VMD \cite{HUMP96, VARSH1994}. The cited algorithm works by counting random points on the surface of each atom, treated as spheres of radius 4.1\AA\ for Si and 4.17\AA\ for Sn, and then deleting those points that fall within their neighbors.

\section{Results and discussion}

\subsection{DFTB parametrization}

The DFTB parameterization shown in this work implies that the main properties of the reference systems are well reproduced with respect to the DFT. This can be achieved by fitting the wave function and density confinement radii for Sn and Si, as well as by accurately constructing repulsion profiles for all diatomic cases of Sn-Sn, Sn-Si, and Si-Si.

\begin{table}
\centering
\begin{tabular}{lllllllll}
\hline
element & $r_0.\phi$ & $\sigma.\phi$ & $r_0.\rho^0$ & $\sigma.\rho^0$& $\epsilon_s$ & $\epsilon_p$ & $U_s$ & $U_p$\\
\hline
  Si     &$4.318$ &$5.815$ & $6.317$&$2.139$ & $-0.395$&$-0.150$ &$0.329$ & $0.244$\\
  Sn     &$5.648$ &$2.000$ & $5.648$&$2.000$ & $-0.360$&$-0.138$ &$0.291$ & $0.215$\\
\hline
\end{tabular}
\caption{Confinement potential parameters $r_0$ (Bohr) and $\sigma$, for atomic orbitals $\phi$ and densities $\rho^0$ of Si and Sn.}
\label{vconf_param}
\end{table}

In our previous DFTB parameterization of the Si-Li system~\cite{oviedo_2023}, we introduce a new electronic structure optimization procedure that weights the chemical structures differently to improve the prediction of their formation energies. The model resulting from the application of this new protocol proved to be robust in predicting crystalline and amorphous structures for the entire range of Si-Li compositions, showing excellent agreement with DFT calculations and exceeding the state-of-the-art ReaxFF potentials. First, the empirical parameter $V_{\mathrm{conf}}$ (equation \ref{vconf}) is optimized using the reference DFT with the methodology implemented in the Hotcent code \cite{vandenbossche_2019a, vandenbossche_2019}. Table \ref{vconf_param} shows the confinement parameters $r_0$ and $s$ obtained by fitting the band structure of Si (mp-149) and Sn (mp-117) separately to the DFT reference. We have considered only the 3s and 3p valence electrons for Si and the 4s and 4p for Sn. The resulting band structures of Sn are shown in Figure S1 of the Supporting Information. Note that for the case of Si we use the same parameters as in our previous work~\cite{oviedo_2023} and the corresponding band structures can be seen in Figure S1 of that reference.

\begin{figure}
\center\includegraphics[width=\linewidth]{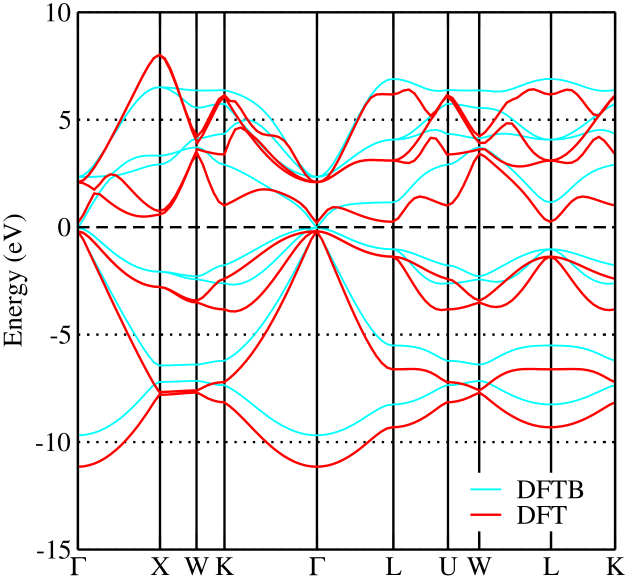}
\caption{Band structure computed by DFTB, in comparison with the band structures computed by DFT/PBE for SiSn. The electronic bands are shifted to the respective Fermi levels (0 eV).
}
\label{bands}
\end{figure}
Figure \ref{bands} shows a comparison of the DFTB and DFT band structures for the SiSn system with a zincblende crystal structure with point group symmetry $F\bar{4}3m$, lattice constant of $6.07$ \AA, as reported in the materials project (mp-1009813)~\cite{jain_2013}. The resulting band structures have been shifted so that the Fermi energy of each structure coincides with the origin of the energy axis. Overall, for the bulk SiSn system, we find that most of the bands are compressed compared to the DFT analogues, but the shape of the bands is preserved. In addition, the DFTB band structure closely resembles the DFT bands near the Fermi energy level. Note that the DFTB parameters were fitted using the single element crystals as mentioned above and not using this SiSn crystal structure. Therefore, the present results demonstrate the excellent predictive power of the developed DFTB model.

To complete the model the $V_{\mathrm{rep}}$ (equation \ref{v_rep}) parameters are optimized using the ``Milonga'' algorithm presented in our recent article~\cite{oviedo_2023}, which uses the TANGO code \cite{vandenbossche_2018} with the aim of predicting relative formation energies of different compositions. As we did in our previous work, we build the training data set using compressions and expansions of the bulk structures of pure Si (mp-149), Sn (mp-117)  and the metastable SiSn (mp-1009813)~\cite{zhang_2012a}. The DFTB model obtained is able to reproduce the formation energy of the SiSn structure with a difference of only $0.0026$ eV below the reference value calculated with DFT. 

\begin{figure}
\center\includegraphics[width=\linewidth]{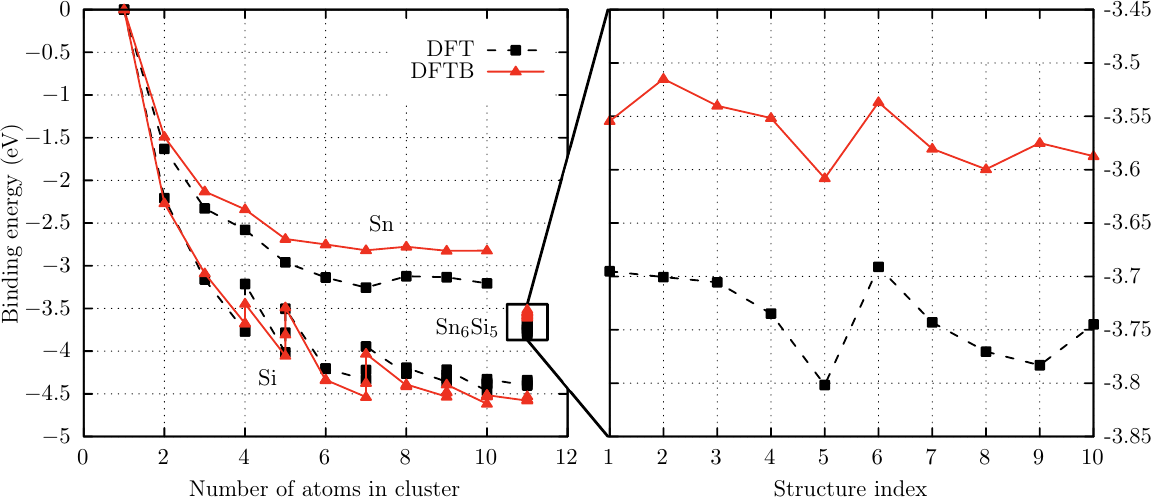}
\caption{Comparison of binding energies computed with DFT and the new DFTB model for a series of Sn$_n$ and Si$_n$ nanoparticles, with $n=1,\cdots,11$, and also for different conformations of the Sn$_6$Si$_5$ nanoparticle taken from MD simulation.}
\label{nano}
\end{figure}

To investigate the predictive power of the new DFTB model, we use it to compute the binding energy of the Sn$_n$ and Si$_n$ nanoparticles (NPs). The Si NPs structures were taken from \cite{fernandes2018quantitative} and Sn were obtained by radial sectioning of the bulk structure. A comparison of the resulting energies with DFT can be seen in figure~\ref{nano}. Then, we used the new DFTB model to perform a MD simulation of the collision of Sn$_6$ and Si$_5$ NPs to prepare an Sn$_6$Si$_5$ bimetallic one. Then we took 10 different conformations at different simulation times after the collision and compute also their binding energy. The results are shown in Figure~\ref{nano}b. Although a small difference of about $\sim$0.15eV is found with respect to DFT, the relative tendency between different nanoparticle conformations is adequately reproduced. These promising results indicate that the new DFTB model, trained on bulk structures, is highly transferable to nanoscopic systems.

\subsection{Collision dynamics for the case of small Si particles}

\begin{figure}
\center\includegraphics[width=\linewidth]{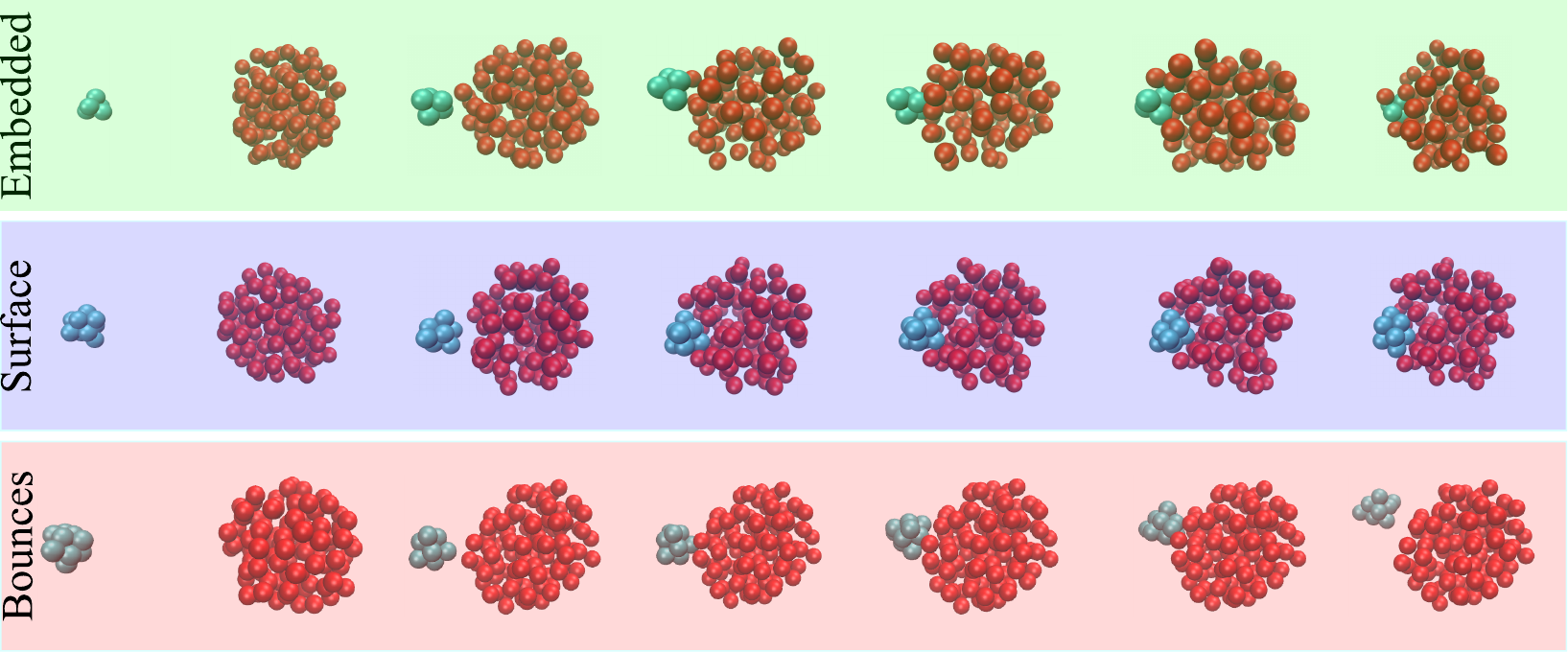}
\caption{Three different regimes found after the collision: embedding, surface-surface interaction and bounce off. Si and Sn atoms are represented as cyan and red particles, respectively.}
\label{conf}
\end{figure}

The collision of the smaller Si clusters (N$_{\mathrm{Si}}$=5, 10 and 11) against the Sn nanoparticle was found to develop in three different regimes during the simulation, Figure \ref{conf}. Si and Sn atoms are represented in cyan and red, respectively. The regimes pictured in Figure \ref{conf} were summarized as:
\begin{enumerate}
\item The Si particle gets embedded into the Sn surface (green box).
\item The particle hits the surface and stays in the same place or moves over the surface of the Sn cluster (blue box).
\item The Si particle impacts on the Sn surface and bounces off (red box).
\end{enumerate}
Theoretical investigations using MD, Monte Carlo and DFT \cite{eom2021general,wang2009predicted} have shown that the cohesive energy ($\Delta$E$_c$) is the predominant descriptor, along with the Wigner-Seitz radius (WS$_r$), to discriminate when a metal prefers to occupies the core or the shell. The larger the $\Delta$E$_c$ and the smaller the WS$_r$, the greater the trend that the metal tends to form the part of the core. The cohesive energies of silicon and tin are $\Delta$E$_c$=4.63 eV/atom and $\Delta$E$_c$=3.14 eV/atom \cite{kittel2005introduction}, respectively. Furthermore, the WS$_r$ of Si is smaller than the Sn WS$_r$. Thus, according to this picture, we expect that Si and Sn could yield a core-shell structure with Si in the core, similar to the ``embedded'' regime shown in figure~\ref{conf}. However, our simulations reveal a much more complex behavior, including  other collision outcomes such as bouncing or surface diffusion of the Si particle.

\begin{figure}
\center
{\includegraphics[width=\linewidth]{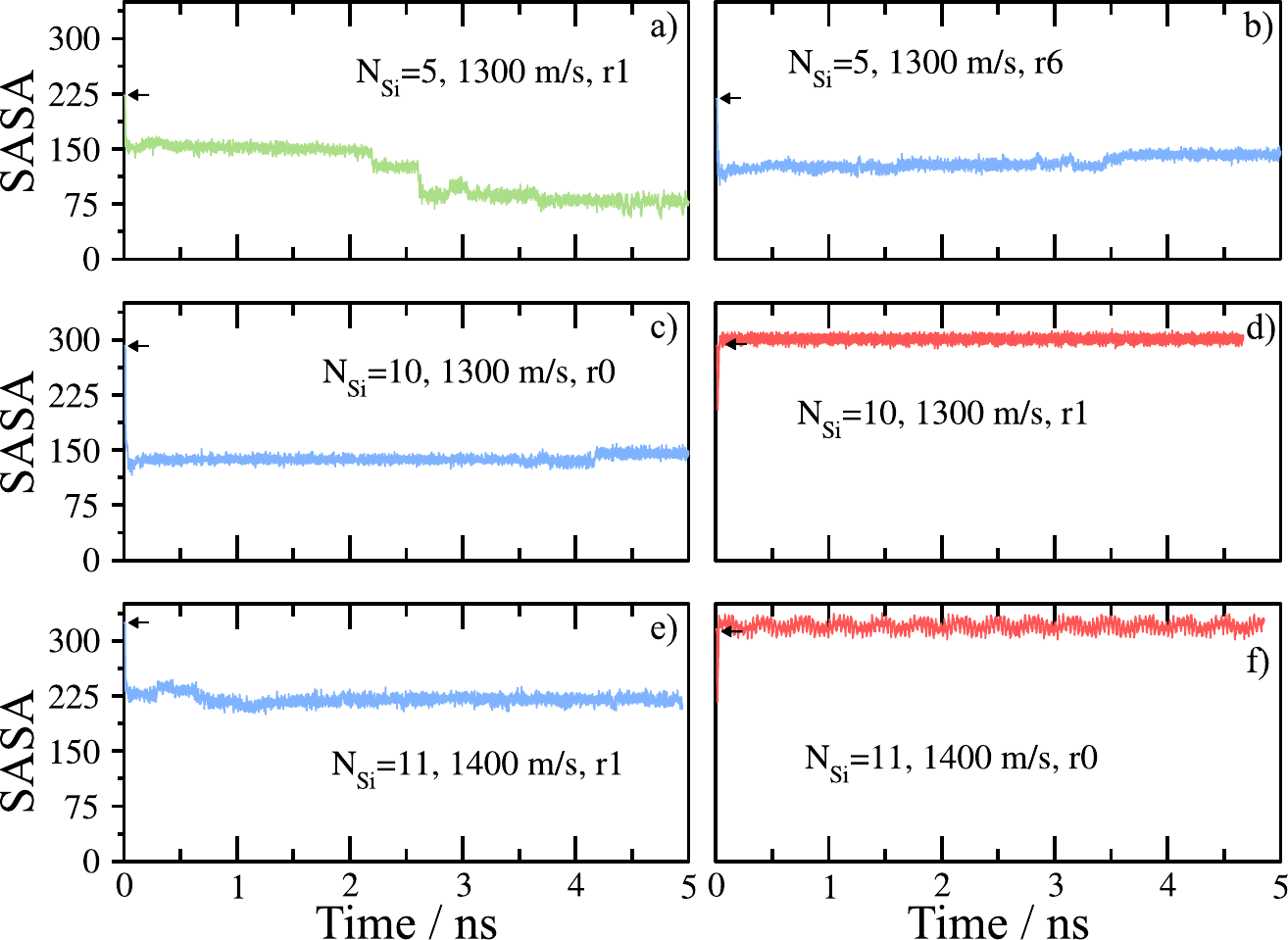}}
\caption{SASA of collisions of Si clusters with different numbers of atoms against a Sn cluster. The SASA gives an insight into collision evolution. The arrow  in each plot shows the SASA at t=0}
\label{reg}
\end{figure}

In order to quantify the extent to which the Si particle is buried into the Sn particle, we have computed a property commonly known as solvent-accessible surface area (SASA) for the Si atoms. Although no solvent is explicitly included in our simulations, this quantity allows to measure the exposure of Si surface to the external environment. For simplicity, three cases were chosen to illustrate the general behaviours. Figure \ref{reg}a shows a typical case of a Si$_5$ cluster being embedded in the Sn particle. At the very beginning the SASA trace shows a sudden drop from its initial value, indicated by a left arrow in the plot. This drop is associated with the initial contact of the two particles, given the significant reduction of the surface exposure of the Si nanoparticle. The evolution of SASA is then observed in stages, with a succession of steady states and sudden changes, which is characteristic of rare events. In this case, the rare events generally tend to decrease the SASA trace, so we can associate them with the gradual burial of the Si particle in the Sn particles. This complex process is shown to involve several steps, certainly including many Sn-Sn bond cleavage and Sn-Si bond formation. We considered that the particle is embedded or mostly buried when the final SASA is less than half of the initial SASA.

Figure \ref{reg}b shows the SASA of another typical case, when Si$_5$ particle collides with the Sn$_{110}$ cluster and remains on the surface. In this case, the SASA suffer the initial drop but then reaches a steady value. This behaviour, also observed in Figure \ref{reg}c and e for Si$_{10}$ and Si$_{11}$, reflects the fact that the Si particle impacts on the Sn surface and is not able to break the Sn atoms bonds and penetrate into the Sn particle. The small fluctuations in the plateau value are associated with rotations of the Si cluster hiding or exposing more surface area.

Figure \ref{reg}d (Si$_{10}$) and f (Si$_{11}$) show a very interesting behaviour. After the initial drop of the SASA, the trace returns to its initial value and starts to oscillate. These conditions are associated with the bouncing off of the Si cluster on the Sn surface. The  oscillatory behavior is produced by the collective harmonic oscillation of the Si atoms which reduces and expands the Si surface. 

\begin{figure}
\center
{\includegraphics[width=\linewidth]{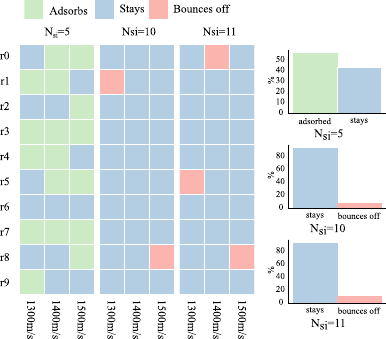}}
\caption{Graphical matrix of the collisions, each color represents to which state the system evolves. The different simulations performed are labelled from r0 to r9.}
\label{mat}
\end{figure}

Summarising, the behaviours found are: the SASA first drops abruptly, remains steady and drops again. The final SASA values are located below half of the initial ones. This means that the Si particle became partially embedded; the SASA abruptly falls and then increases or reduces, always well below the initial value. The final values are located above half of the initial SASA. This means the Si particle stays on the Sn surface. We will denominate this situation as a ``contact'' case and the SASA decays and increases again to an oscillating value around the initial SASA. This means the Si particle is bounced from the Sn particle. The SASA regime landscape can be easily represented by the graphical matrix shown in Figure~\ref{mat}, where each simulation performed is highlighted according to its collision outcome. For Si$_5$, more than 50\%  of the collisions result in the Si cluster being embedded into the Sn particle. There is a significant difference for Si$_{10}$ and Si$_{11}$. In these cases,  most of the collisions end with the Si cluster going around the Sn surface. The cases where Si bounces off represent around 10\% of the cases. 

\begin{figure}
\center
{\includegraphics[width=\linewidth]{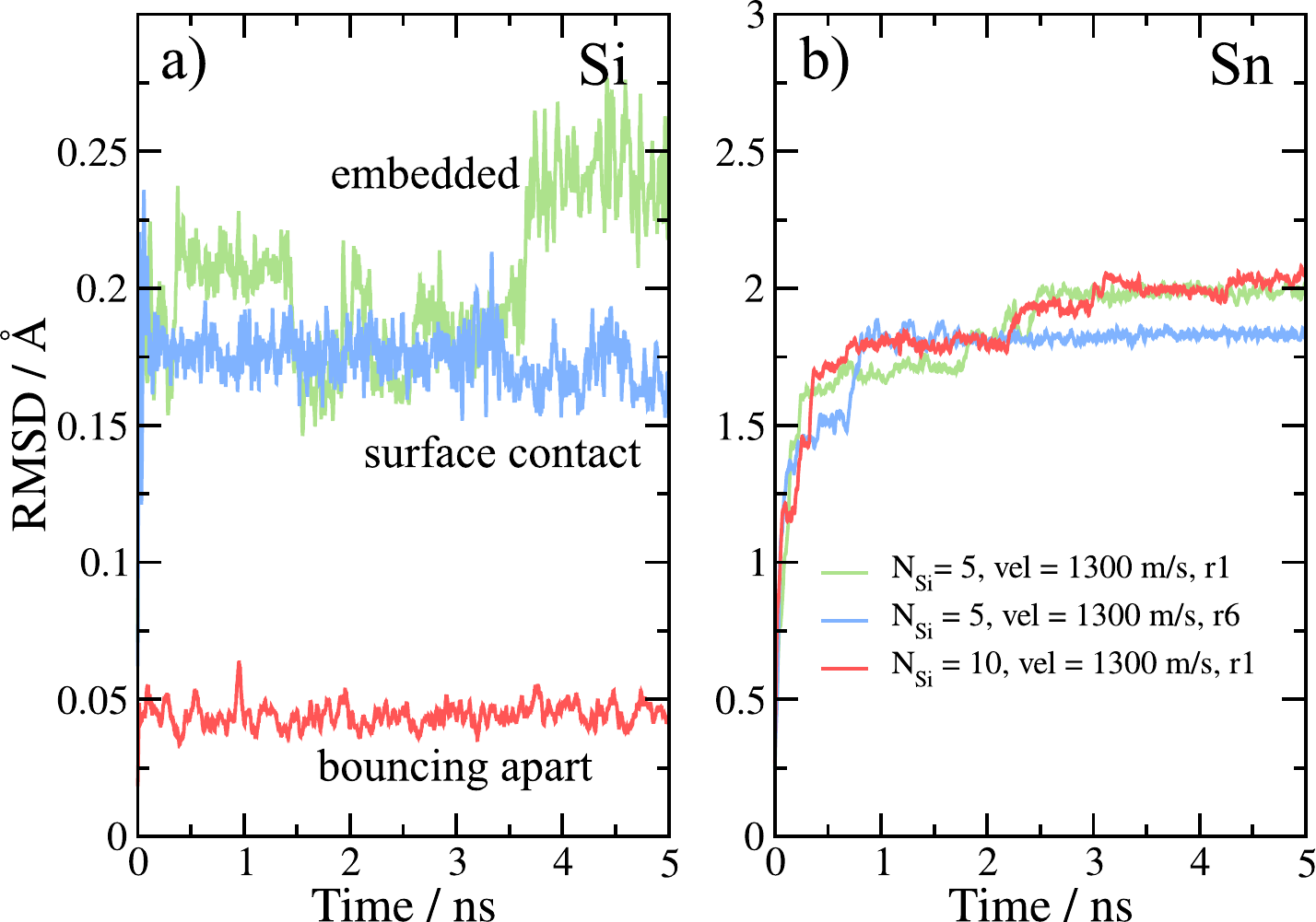}}
\caption{RMSD, for the systems analyzed in Fig. \ref{reg}. Depending on the regime the RMSD for Si reaches a small steady state value or increases.The color key corresponds to the different behaviors shown in Figure~\ref{conf}: green is the Si embedded in Sn, blue is the particles in surface contact, and red is their bouncing apart.}
\label{def}
\end{figure}

We now turn to analyze the degree of deformation of each nanoparticle during the collision. To do this, the nanoparticle conformation of each simulation frame was centered and aligned to match the initial conformation before the collision. Then we compute the Root Mean Square Distance (RMSD), as shown in figure~\ref{def}.

To illustrate the RMSD behaviour three representative systems were chosen, at the same impact velocity, for a Si particle being embedded, staying at the surface and bouncing apart. Figure~\ref{def}a shows the RMSD for Si for the systems chosen. The green curve corresponds to a Si$_5$ (1300 m/s, r1) particle that penetrates into the Sn surface. In this case, the Si deformation parameter oscillates first around 0.2 \AA\ and jumps to 0.25 \AA\ as the Si$_5$ particle becomes buried into the Sn$_{110}$ one. This occurs around 0.35 ns, and the Si particle becomes compressed. The blue curve shows the RMSD of another Si$_5$ (1300 m/s, r6) particle that does not penetrate into the Sn surface but remains on it. The RMSD shows an abrupt increase and then fluctuates around 0.18 \AA\. Finally, the red curve shows a Si$_{10}$ particle (1300 m/s, r1) that bounces off. The Si RMSD shows a small deformation due to the fact that  Si is no longer interacting with the Sn$_{110}$ particle. Fig. \ref{def}b shows the RMSD for Sn for the same systems mentioned before. In the three cases, the Sn deformation shows a similar behaviour as it increases after impact, but it reaches an immediate steady state only when the Si remains on its surface (blue curve). When the Si particle bounces back (red curve) the Sn jumps between states of greater deformation. Figure S3 in Supporting Information shows all the Si RMSD, its possible to observe that after the collision, the number of collisions in which the RMSD reaches larger values increases as the Si particle gets larger. 

To analyze the particles' energetics, the kinetic energy per element was calculated using the instantaneous velocity of the atoms, relative to the velocity of the center of mass of each nanoparticle:
\begin{equation}
    \mathrm{E}_K^{\mathrm{i}}=\dfrac{m^{\mathrm{i}}}{2}\sum_j(v_j^\mathrm{i}(t)-v_{cm}^\mathrm{i}(t))^{2},
\label{eq1}
\end{equation}
with i=Si, Sn. Then we define the relative kinetic energy RE$_K$ as:
\begin{equation}
    \mathrm{RE}_K = \dfrac{\mathrm{E}_K^{\mathrm{Sn}}}{\mathrm{E}_K^{\mathrm{Si}}}.
\label{eq2}
\end{equation}

\begin{figure}
\center
{\includegraphics[width=\linewidth]{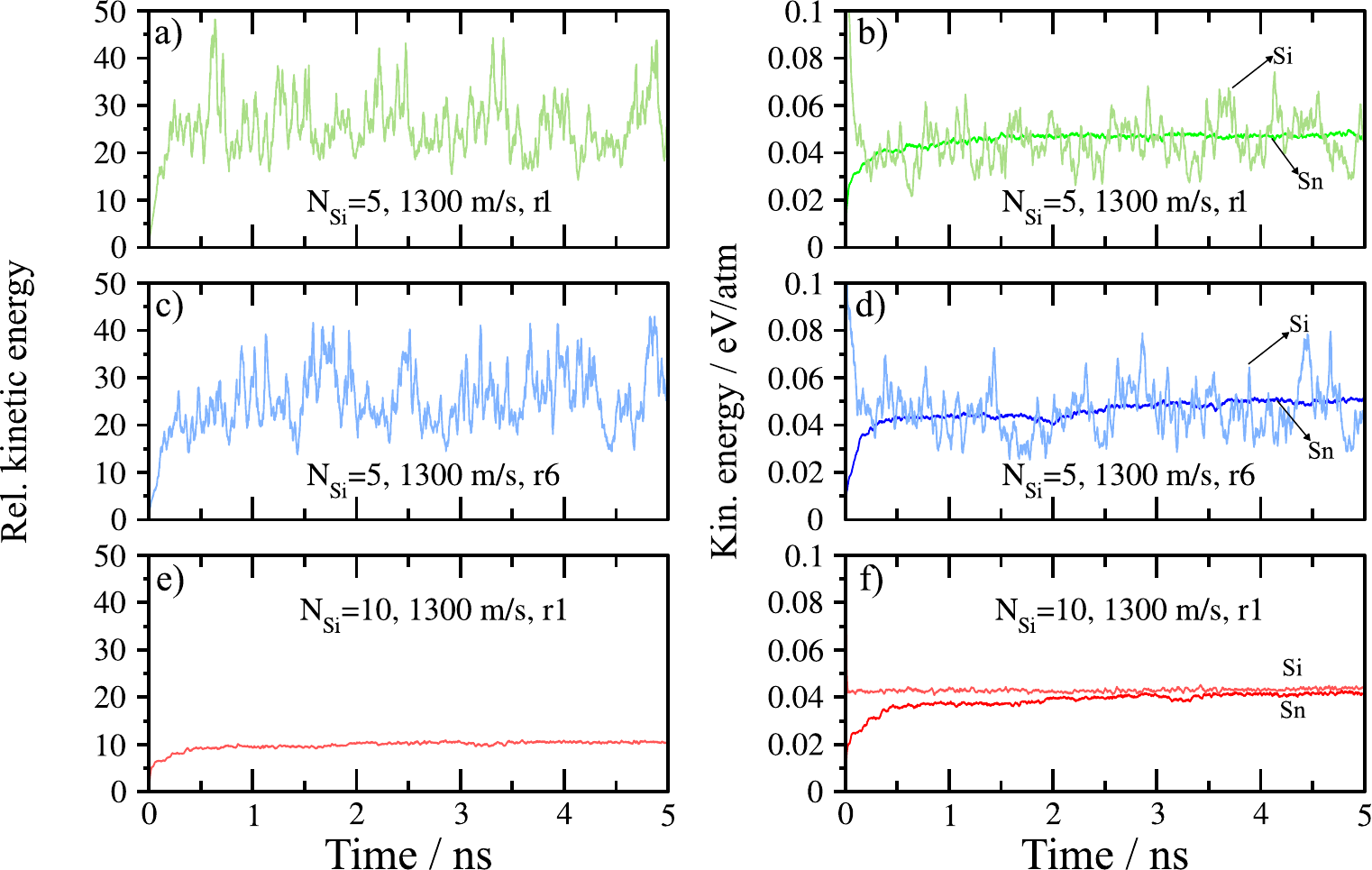}}
\caption{Left: Relative kinetic energy between Sn and Si clusters in both colliding clusters, as defined in equation (8). Right:  kinetic energy per element for the collisions of a cluster with N$_{\mathrm{Si}}=5$ 1400 m/s, r2 and a cluster with N$_{\mathrm{Si}}=10$ 1300 m/s, r1. The color key corresponds to the different behaviors shown in Figure~\ref{conf}: green is the Si embedded in Sn, blue is the particles in surface contact, and red is their bouncing apart.}
\label{ken}
\end{figure}

Figure~\ref{ken} shows E$_K^{\mathrm{Si,Sn}}$ and RE$_K$ of three selected simulations, at the same impact velocity, which are chosen to represent the three typical cases found in this work: Si is embedded in the Sn, the particles remain in surface contact, or they bounce apart. The green curve, Figure~\ref{ken}a and b, shows the kinetic energy when the Si$_5$ (1300 m/s r1) particle collides with the Sn$_{110}$ and the Si is embedded in the Sn tending to form a core-shell-like structure. If we consider that the mean energy per particle tends to 3$k_B$T/2, RE$_K$ should tend to N$_{Sn}$/N$_{Si}=22$. Figure~\ref{ken}a shows strong fluctuations of RE$_K$ around this value. These fluctuations are due to the fact that the average velocity of the Si atoms changes abruptly at each instant. A detailed analysis of the velocity per element, Fig. \ref{ken}b, shows that after impact the Si particle delivers kinetic energy to tin. While E$_K^{\mathrm{Sn}}$ increases steadily, E$_K^{\mathrm{Si}}$ shows fluctuations around the asymptotic value of E$_K^{\mathrm{Sn}}\sim0.05$ eV/atom. Fig. \ref{ken}c and d, blue curves, show RE$_K$, E$_K^{\mathrm{Sn}}$ and E$_K^{\mathrm{Si}}$ for a system in which a Si$_5$ (1300 m/s r6) cluster collides with Sn and remains at the tin surface. The energetics, in this case, resembles the result observed in Fig. \ref{ken}a and b, but E$_K^{\mathrm{Sn}}$ tends to be around 0.047 eV/atom. A different picture arises when the Si$_{10}$ (1300 m/s r1) bounces off, red curves. RE$_K$ seems to reach a steady state around 110/10=11, meaning that the energy is equally distributed in both clusters. E$_K^{\mathrm{Sn}}$ and E$_K^{\mathrm{Si}}$ show that after 2 ns the system reaches an equilibrium with an average E$_K^{\mathrm{Sn}}$ and E$_K^{\mathrm{Si}}$ close to 0.042 eV/atom.

\begin{figure}
 \center
 {\includegraphics[width=\linewidth]{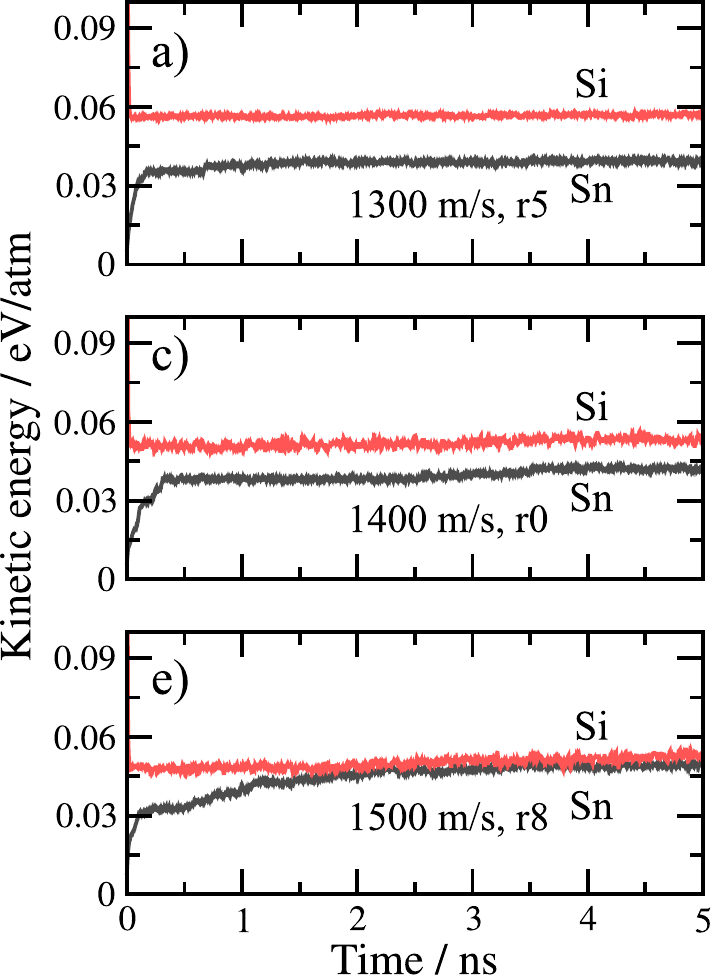}}
 \caption{a): Kinetic energy per atom for the collisions in which the Si$_{11}$ bounces off: 1300 m/s r5, 1400 m/s r0, 1500 m/s r8. As collision velocity increases, the gap between E$_K^{\mathrm{Si}}$ and E$_K^{\mathrm{Sn}}$ decreases.}
 \label{ken2}
 \end{figure}

We will compare now the E$_K^{\mathrm{Si,Sn}}$ for three simulations were Si$_{11}$ particle collides at 1300 m/s (r5), 1400 m/s (r0) and 1500 m/s (r8) and bounces apart. Figures \ref{ken2}a, c and e show that as collision velocity increases, the gap between E$_K^{\mathrm{Si}}$ and E$_K^{\mathrm{Sn}}$ decreases, indicating that a stronger collision seems to favor the energy exchange between the particles. At 1300 m/s the average Si kinetic energy per atom is around 20\% greater than the Sn energy, while at 1500 m/s the collision transfers enough energy to reach the same energy for each atom. The fact that energy equilibration improves with higher initial velocities also correlates with a longer period of time that the particle stays in contact, as can be seen by the downward arrows indicating the moment of particle separation.

\subsection{Collision dynamics for the case of comparable particle sizes}

\begin{figure}
\center
{\includegraphics[width=\linewidth]{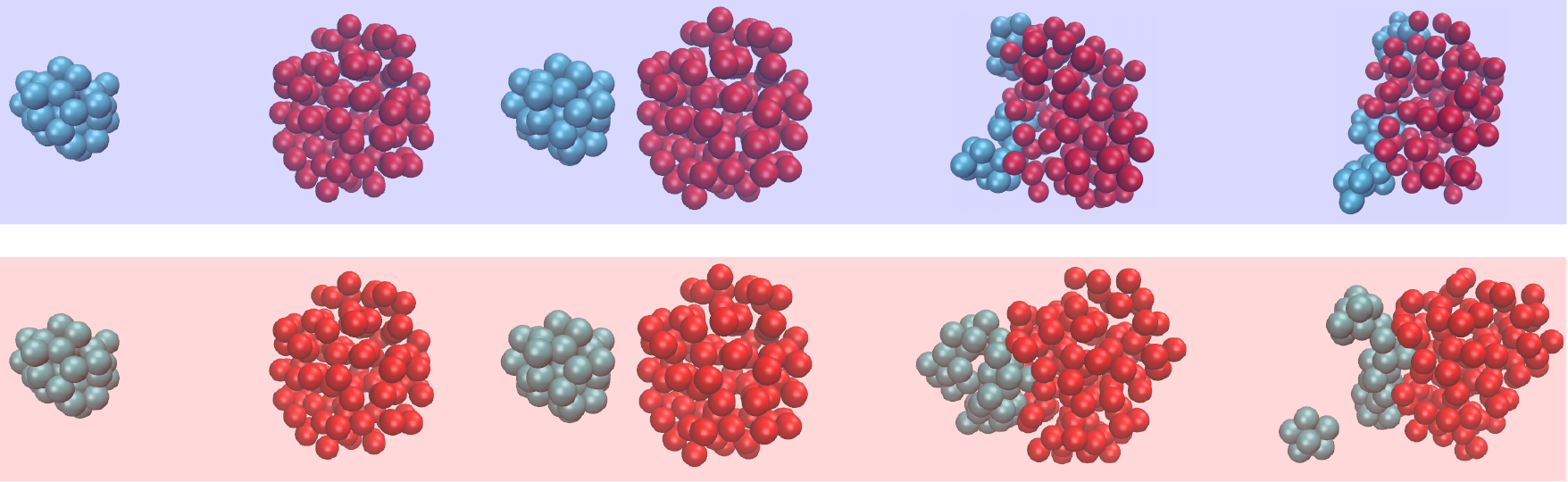}}
\caption{Representative snapshots for the collisions between Si$_{40}$ and Sn$_{110}$. In all cases the Si particle is broken into smaller clusters.}
\label{reg40}
\end{figure}

A new scenario arises as the Si particle becomes bigger. For Si$_{40}$ the collision produces the breakup of the particle into smaller Si clusters. These clusters show an individual behaviour similar to those studied before: they could bounce off, stay at or penetrate the Sn surface. In this case, for each of the three velocities studied before, we perform five simulation with different random states, all of them showing the breaking of Si into small particles after the collision. These simulations can be classified into two distinctive behaviors, as shown in Figure \ref{reg40}. From the 15 simulations, 8 (53\%) evolve to a situation in which the small Si clusters formed stay at the surface of the Sn nanoparticle, see Figure S1a in the supporting information. The other 7 systems (43\%) show at least one cluster bouncing away. 

\begin{figure}
\center\includegraphics[width=\linewidth]{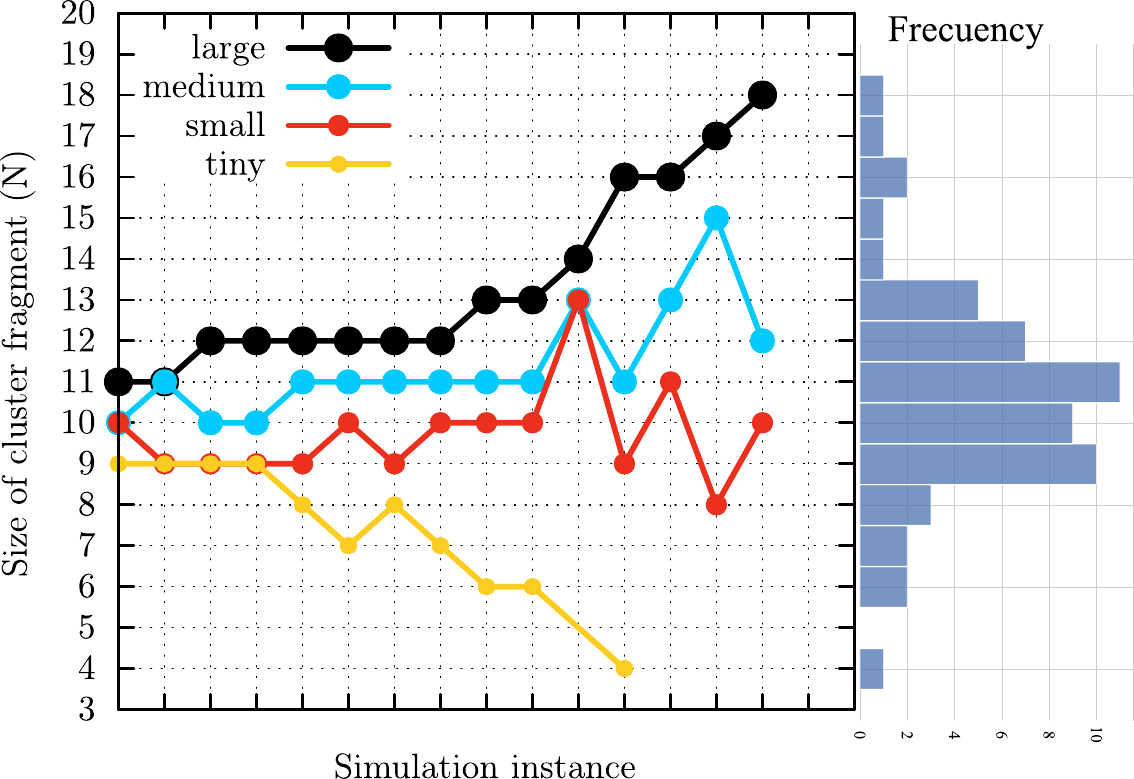}
\caption{Si cluster size distribution after impact. N$_{\mathrm{Si}}=40$  }
\label{mat40}
\end{figure}

We will first focus into characterize the size distribution of the clusters formed during the Si particle rupture. In the general case, the Si particle breaks into clusters with sizes that vary between 9 and 13 atoms each, as can be noted by the peak in the histogram of Figure \ref{mat40}b. However, since the sum of all Si atoms must be always 40, there is a strong correlation between the size of the different cluster formed. This can be seen in Figure~\ref{mat40}a, which shows the detailed size and number of clusters obtained for each of the 15 simulation instances. The data are sorted on the horizontal axes, and the points are connected by a line that serves as a guide to the eye to facilitate observation of the size correlation. Note that for each simulation, 3 or 4 different clusters are formed after the Si particle fracture, with 4 being the most common number (80\%). We have labelled these 4 clusters as tiny, small, medium and large. It is interesting to note that the large cluster increases its size at the expense of the tiny one and the systems seem to favor the small and medium sizes between 9 and 12. According to Fernandes the most stable Si clusters are those with 10, 7, 9 and 8 atoms respectively \cite{fernandes2018quantitative}. Of the 56 clusters, 24 have sizes in the above range and the 14 clusters have unstable sizes. The other 18 clusters have sizes beyond the analysis of the mentioned work.

\section{Conclusions}

We have developed a DFTB framework to model Si-Sn systems and applied it to study the collisions between two pure clusters to form a bimetallic one. This system is of interest for applications of Si/Sn composites in Li-ion batteries, since in some of the developments the composites are prepared by a ball-milling process, which involves the high-energy collision between Si and Sn particles. Although a core-shell Si@Sn structure can be expected as the collision product by considering the binding energies of the materials, our simulations show a much more interesting scenario. When a small Si cluster collides with a Sn particle, a variety of outcomes are possible, such as the particles bouncing apart, remaining in surface contact, or the Si particle being buried. For the case of Si particles of larger size, the fragmentation into small clusters is found, each of them exhibiting the same behaviors described before. We claim that the diversity and complexity revealed by our simulations is only possible thanks to the new Sn-Si DFTB model developed here. Although we have considered relatively small particles here, our results motivate further research on collision dynamics in this and other systems using DFTB.

\section{Acknowledgement}
This work used computational resources from CCAD-UNC, which is part of
SNCAD-MinCyT, Argentina.  We acknowledge financial support from CONICET
(28720210100623CO, 28720210101190CO, 1220200101189CO, PUE/2017), the Agencia
Nacional de Promoci\'on Cient\'ifica y Tecnol\'ogica (FONCyT 2020-SERIEA-02139,
2020-SERIEA-03689) and SECyT of the Universidad Nacional de C\'ordoba.

\bibliography{manuscript_sisn}

\providecommand{\latin}[1]{#1}
\makeatletter
\providecommand{\doi}
  {\begingroup\let\do\@makeother\dospecials
  \catcode`\{=1 \catcode`\}=2 \doi@aux}
\providecommand{\doi@aux}[1]{\endgroup\texttt{#1}}
\makeatother
\providecommand*\mcitethebibliography{\thebibliography}
\csname @ifundefined\endcsname{endmcitethebibliography}
  {\let\endmcitethebibliography\endthebibliography}{}
\begin{mcitethebibliography}{70}
\providecommand*\natexlab[1]{#1}
\providecommand*\mciteSetBstSublistMode[1]{}
\providecommand*\mciteSetBstMaxWidthForm[2]{}
\providecommand*\mciteBstWouldAddEndPuncttrue
  {\def\EndOfBibitem{\unskip.}}
\providecommand*\mciteBstWouldAddEndPunctfalse
  {\let\EndOfBibitem\relax}
\providecommand*\mciteSetBstMidEndSepPunct[3]{}
\providecommand*\mciteSetBstSublistLabelBeginEnd[3]{}
\providecommand*\EndOfBibitem{}
\mciteSetBstSublistMode{f}
\mciteSetBstMaxWidthForm{subitem}{(\alph{mcitesubitemcount})}
\mciteSetBstSublistLabelBeginEnd
  {\mcitemaxwidthsubitemform\space}
  {\relax}
  {\relax}

\bibitem[Li \latin{et~al.}(2018)Li, Levine, Zhang, Lee, Naskar, Dai, and
  Paranthaman]{li2018carbon}
Li,~Y.; Levine,~A.~M.; Zhang,~J.; Lee,~R.~J.; Naskar,~A.~K.; Dai,~S.;
  Paranthaman,~M.~P. Carbon/tin oxide composite electrodes for improved
  lithium-ion batteries. \emph{Journal of Applied Electrochemistry}
  \textbf{2018}, \emph{48}, 811--817\relax
\mciteBstWouldAddEndPuncttrue
\mciteSetBstMidEndSepPunct{\mcitedefaultmidpunct}
{\mcitedefaultendpunct}{\mcitedefaultseppunct}\relax
\EndOfBibitem
\bibitem[Xu \latin{et~al.}(2018)Xu, Sun, Yin, and Guo]{xu2018facile}
Xu,~Q.; Sun,~J.-K.; Yin,~Y.-X.; Guo,~Y.-G. Facile synthesis of blocky SiOx/C
  with graphite-like structure for high-performance lithium-ion battery anodes.
  \emph{Advanced Functional Materials} \textbf{2018}, \emph{28}, 1705235\relax
\mciteBstWouldAddEndPuncttrue
\mciteSetBstMidEndSepPunct{\mcitedefaultmidpunct}
{\mcitedefaultendpunct}{\mcitedefaultseppunct}\relax
\EndOfBibitem
\bibitem[Smrekar \latin{et~al.}(2020)Smrekar, Bracamonte, Primo, Luque, Thomas,
  Barraco, and Leiva]{smrekar2020mapping}
Smrekar,~S.; Bracamonte,~M.~V.; Primo,~E.~N.; Luque,~G.~L.; Thomas,~J.;
  Barraco,~D.~E.; Leiva,~E. A Mapping of the Physical and Electrochemical
  Properties of Composite Lithium-Ion Batteries Anodes Made from Graphite, Sn,
  and Si. \emph{Batteries \& Supercaps} \textbf{2020}, \emph{3},
  1248--1256\relax
\mciteBstWouldAddEndPuncttrue
\mciteSetBstMidEndSepPunct{\mcitedefaultmidpunct}
{\mcitedefaultendpunct}{\mcitedefaultseppunct}\relax
\EndOfBibitem
\bibitem[Ruderman \latin{et~al.}(2021)Ruderman, Smrekar, Bracamonte, Primo,
  Luque, Thomas, Leiva, Monti, Barraco, and Ch{\'a}vez]{ruderman2021unveiling}
Ruderman,~A.; Smrekar,~S.; Bracamonte,~M.~V.; Primo,~E.~N.; Luque,~G.~L.;
  Thomas,~J.; Leiva,~E.; Monti,~G.~A.; Barraco,~D.~E.; Ch{\'a}vez,~F.~V.
  Unveiling the stability of Sn/Si/graphite composites for Li-ion storage by
  physical, electrochemical and computational tools. \emph{Physical Chemistry
  Chemical Physics} \textbf{2021}, \emph{23}, 3281--3289\relax
\mciteBstWouldAddEndPuncttrue
\mciteSetBstMidEndSepPunct{\mcitedefaultmidpunct}
{\mcitedefaultendpunct}{\mcitedefaultseppunct}\relax
\EndOfBibitem
\bibitem[Elemike \latin{et~al.}(2019)Elemike, Onwudiwe, Nundkumar, Singh, and
  Iyekowa]{elemike_2019}
Elemike,~E.~E.; Onwudiwe,~D.~C.; Nundkumar,~N.; Singh,~M.; Iyekowa,~O. Green
  synthesis of {Ag}, {Au} and {Ag}-{Au} bimetallic nanoparticles using
  {Stigmaphyllon} ovatum leaf extract and their in vitro anticancer potential.
  \emph{Materials Letters} \textbf{2019}, \emph{243}, 148--152\relax
\mciteBstWouldAddEndPuncttrue
\mciteSetBstMidEndSepPunct{\mcitedefaultmidpunct}
{\mcitedefaultendpunct}{\mcitedefaultseppunct}\relax
\EndOfBibitem
\bibitem[Lomelí-Marroquín \latin{et~al.}(2019)Lomelí-Marroquín, Cruz,
  Nieto-Argüello, Crua, Chen, Torres-Castro, Webster, and
  Cholula-Díaz]{lomeli-marroquin_2019}
Lomelí-Marroquín,~D.; Cruz,~D.~M.; Nieto-Argüello,~A.; Crua,~A.~V.;
  Chen,~J.; Torres-Castro,~A.; Webster,~T.~J.; Cholula-Díaz,~J.~L.
  {\textless}p{\textgreater}{Starch}-mediated synthesis of mono- and bimetallic
  silver/gold nanoparticles as antimicrobial and anticancer
  agents{\textless}/p{\textgreater}. \emph{International Journal of
  Nanomedicine} \textbf{2019}, \emph{14}, 2171--2190, Publisher: Dove
  Press\relax
\mciteBstWouldAddEndPuncttrue
\mciteSetBstMidEndSepPunct{\mcitedefaultmidpunct}
{\mcitedefaultendpunct}{\mcitedefaultseppunct}\relax
\EndOfBibitem
\bibitem[Sivamaruthi \latin{et~al.}(2019)Sivamaruthi, Ramkumar, Archunan,
  Chaiyasut, and Suganthy]{sivamaruthi_2019}
Sivamaruthi,~B.~S.; Ramkumar,~V.~S.; Archunan,~G.; Chaiyasut,~C.; Suganthy,~N.
  Biogenic synthesis of silver palladium bimetallic nanoparticles from fruit
  extract of {Terminalia} chebula – {In} vitro evaluation of anticancer and
  antimicrobial activity. \emph{Journal of Drug Delivery Science and
  Technology} \textbf{2019}, \emph{51}, 139--151\relax
\mciteBstWouldAddEndPuncttrue
\mciteSetBstMidEndSepPunct{\mcitedefaultmidpunct}
{\mcitedefaultendpunct}{\mcitedefaultseppunct}\relax
\EndOfBibitem
\bibitem[Minal and Prakash(2020)Minal, and Prakash]{minal_2020}
Minal,~S.~P.; Prakash,~S. Laboratory analysis of {Au}–{Pd} bimetallic
  nanoparticles synthesized with {Citrus} limon leaf extract and its efficacy
  on mosquito larvae and non-target organisms. \emph{Scientific Reports}
  \textbf{2020}, \emph{10}, 21610\relax
\mciteBstWouldAddEndPuncttrue
\mciteSetBstMidEndSepPunct{\mcitedefaultmidpunct}
{\mcitedefaultendpunct}{\mcitedefaultseppunct}\relax
\EndOfBibitem
\bibitem[Arora \latin{et~al.}(2020)Arora, Thangavelu, and
  Karanikolos]{arora_2020}
Arora,~N.; Thangavelu,~K.; Karanikolos,~G.~N. Bimetallic {Nanoparticles} for
  {Antimicrobial} {Applications}. \emph{Frontiers in Chemistry} \textbf{2020},
  \emph{8}, 412\relax
\mciteBstWouldAddEndPuncttrue
\mciteSetBstMidEndSepPunct{\mcitedefaultmidpunct}
{\mcitedefaultendpunct}{\mcitedefaultseppunct}\relax
\EndOfBibitem
\bibitem[Padilla-Cruz \latin{et~al.}(2021)Padilla-Cruz, Garza-Cervantes,
  Vasto-Anzaldo, García-Rivas, León-Buitimea, and
  Morones-Ramírez]{padilla-cruz_2021}
Padilla-Cruz,~A.~L.; Garza-Cervantes,~J.~A.; Vasto-Anzaldo,~X.~G.;
  García-Rivas,~G.; León-Buitimea,~A.; Morones-Ramírez,~J.~R. Synthesis and
  design of {Ag}–{Fe} bimetallic nanoparticles as antimicrobial synergistic
  combination therapies against clinically relevant pathogens. \emph{Scientific
  Reports} \textbf{2021}, \emph{11}, 5351\relax
\mciteBstWouldAddEndPuncttrue
\mciteSetBstMidEndSepPunct{\mcitedefaultmidpunct}
{\mcitedefaultendpunct}{\mcitedefaultseppunct}\relax
\EndOfBibitem
\bibitem[Arif(2023)]{arif_2023}
Arif,~M. A tutorial review on bimetallic nanoparticles loaded in smart organic
  polymer microgels/hydrogels. \emph{Journal of Molecular Liquids}
  \textbf{2023}, \emph{375}, 121346\relax
\mciteBstWouldAddEndPuncttrue
\mciteSetBstMidEndSepPunct{\mcitedefaultmidpunct}
{\mcitedefaultendpunct}{\mcitedefaultseppunct}\relax
\EndOfBibitem
\bibitem[Makada \latin{et~al.}(2023)Makada, Habib, and Singh]{makada_2023}
Makada,~H.; Habib,~S.; Singh,~M. Bimetallic nanoparticles as suitable
  nanocarriers in cancer therapy. \emph{Scientific African} \textbf{2023},
  \emph{20}, e01700\relax
\mciteBstWouldAddEndPuncttrue
\mciteSetBstMidEndSepPunct{\mcitedefaultmidpunct}
{\mcitedefaultendpunct}{\mcitedefaultseppunct}\relax
\EndOfBibitem
\bibitem[Amiripour \latin{et~al.}(2018)Amiripour, Azizi, and
  Ghasemi]{amiripour_2018}
Amiripour,~F.; Azizi,~S.~N.; Ghasemi,~S. Gold-copper bimetallic nanoparticles
  supported on nano {P} zeolite modified carbon paste electrode as an efficient
  electrocatalyst and sensitive sensor for determination of hydrazine.
  \emph{Biosensors and Bioelectronics} \textbf{2018}, \emph{107},
  111--117\relax
\mciteBstWouldAddEndPuncttrue
\mciteSetBstMidEndSepPunct{\mcitedefaultmidpunct}
{\mcitedefaultendpunct}{\mcitedefaultseppunct}\relax
\EndOfBibitem
\bibitem[Ansari \latin{et~al.}(2018)Ansari, Saha, Singha, and Sen]{ansari_2018}
Ansari,~Z.; Saha,~A.; Singha,~S.~S.; Sen,~K. Phytomediated generation of {Ag},
  {CuO} and {Ag}-{Cu} nanoparticles for dimethoate sensing. \emph{Journal of
  Photochemistry and Photobiology A: Chemistry} \textbf{2018}, \emph{367},
  200--211\relax
\mciteBstWouldAddEndPuncttrue
\mciteSetBstMidEndSepPunct{\mcitedefaultmidpunct}
{\mcitedefaultendpunct}{\mcitedefaultseppunct}\relax
\EndOfBibitem
\bibitem[Sohrabi \latin{et~al.}(2022)Sohrabi, Majidi, Asadpour-Zeynali,
  Khataee, and Mokhtarzadeh]{sohrabi_2022}
Sohrabi,~H.; Majidi,~M.~R.; Asadpour-Zeynali,~K.; Khataee,~A.; Mokhtarzadeh,~A.
  Bimetallic {Fe}/{Mn} {MOFs}/{M$\beta$CD}/{AuNPs} stabilized on {MWCNTs} for
  developing a label-free {DNA}-based genosensing bio-assay applied in the
  determination of {Salmonella} typhimurium in milk samples. \emph{Chemosphere}
  \textbf{2022}, \emph{287}, 132373\relax
\mciteBstWouldAddEndPuncttrue
\mciteSetBstMidEndSepPunct{\mcitedefaultmidpunct}
{\mcitedefaultendpunct}{\mcitedefaultseppunct}\relax
\EndOfBibitem
\bibitem[Basavegowda \latin{et~al.}(2020)Basavegowda, Mandal, and
  Baek]{basavegowda_2020}
Basavegowda,~N.; Mandal,~T.~K.; Baek,~K.-H. Bimetallic and {Trimetallic}
  {Nanoparticles} for {Active} {Food} {Packaging} {Applications}: {A} {Review}.
  \emph{Food and Bioprocess Technology} \textbf{2020}, \emph{13}, 30--44\relax
\mciteBstWouldAddEndPuncttrue
\mciteSetBstMidEndSepPunct{\mcitedefaultmidpunct}
{\mcitedefaultendpunct}{\mcitedefaultseppunct}\relax
\EndOfBibitem
\bibitem[Mustieles~Marin \latin{et~al.}(2021)Mustieles~Marin, Asensio, and
  Chaudret]{mustielesmarin_2021}
Mustieles~Marin,~I.; Asensio,~J.~M.; Chaudret,~B. Bimetallic {Nanoparticles}
  {Associating} {Noble} {Metals} and {First}-{Row} {Transition} {Metals} in
  {Catalysis}. \emph{ACS Nano} \textbf{2021}, \emph{15}, 3550--3556\relax
\mciteBstWouldAddEndPuncttrue
\mciteSetBstMidEndSepPunct{\mcitedefaultmidpunct}
{\mcitedefaultendpunct}{\mcitedefaultseppunct}\relax
\EndOfBibitem
\bibitem[Bruno \latin{et~al.}(2023)Bruno, Scuderi, Priolo, Falciola, and
  Mirabella]{bruno_2023}
Bruno,~L.; Scuderi,~M.; Priolo,~F.; Falciola,~L.; Mirabella,~S. Enlightening
  the bimetallic effect of {Au}@{Pd} nanoparticles on {Ni} oxide nanostructures
  with enhanced catalytic activity. \emph{Scientific Reports} \textbf{2023},
  \emph{13}, 3203\relax
\mciteBstWouldAddEndPuncttrue
\mciteSetBstMidEndSepPunct{\mcitedefaultmidpunct}
{\mcitedefaultendpunct}{\mcitedefaultseppunct}\relax
\EndOfBibitem
\bibitem[Tao and Li(2023)Tao, and Li]{tao_2023}
Tao,~F.; Li,~Y. A new type of catalysts: catalysts of singly dispersed
  bimetallic sites. \emph{Trends in Chemistry} \textbf{2023}, \emph{0},
  Publisher: Elsevier\relax
\mciteBstWouldAddEndPuncttrue
\mciteSetBstMidEndSepPunct{\mcitedefaultmidpunct}
{\mcitedefaultendpunct}{\mcitedefaultseppunct}\relax
\EndOfBibitem
\bibitem[Qi \latin{et~al.}(2022)Qi, Yang, Jiang, Han, Wu, Wu, Liu, Wang, and
  Wang]{qi_2022}
Qi,~Y.; Yang,~Z.; Jiang,~Y.; Han,~H.; Wu,~T.; Wu,~L.; Liu,~J.; Wang,~Z.;
  Wang,~F. Platinum {Copper} {Bimetallic} {Nanoparticles} {Supported} on
  {TiO2} as {Catalysts} for {Photo} thermal {Catalytic} {Toluene}
  {Combustion}. \emph{ACS Applied Nano Materials} \textbf{2022}, \emph{5},
  1845--1854, Publisher: American Chemical Society\relax
\mciteBstWouldAddEndPuncttrue
\mciteSetBstMidEndSepPunct{\mcitedefaultmidpunct}
{\mcitedefaultendpunct}{\mcitedefaultseppunct}\relax
\EndOfBibitem
\bibitem[Lee \latin{et~al.}(2020)Lee, Hwang, Lee, Schebarchov, Wy, Grand,
  Auguié, Wi, Cortés, and Han]{lee_2020}
Lee,~S.; Hwang,~H.; Lee,~W.; Schebarchov,~D.; Wy,~Y.; Grand,~J.; Auguié,~B.;
  Wi,~D.~H.; Cortés,~E.; Han,~S.~W. Core–{Shell} {Bimetallic} {Nanoparticle}
  {Trimers} for {Efficient} {Light}-to-{Chemical} {Energy} {Conversion}.
  \emph{ACS Energy Letters} \textbf{2020}, \emph{5}, 3881--3890\relax
\mciteBstWouldAddEndPuncttrue
\mciteSetBstMidEndSepPunct{\mcitedefaultmidpunct}
{\mcitedefaultendpunct}{\mcitedefaultseppunct}\relax
\EndOfBibitem
\bibitem[Bhiradi and Hiremath(2022)Bhiradi, and Hiremath]{bhiradi_2022}
Bhiradi,~I.; Hiremath,~S.~S. Energy storage and photosensitivity of in-situ
  formed silver-copper ({Ag}-{Cu}) heterogeneous nanoparticles generated using
  multi-tool micro electro discharge machining process. \emph{Journal of Alloys
  and Compounds} \textbf{2022}, \emph{897}, 162950\relax
\mciteBstWouldAddEndPuncttrue
\mciteSetBstMidEndSepPunct{\mcitedefaultmidpunct}
{\mcitedefaultendpunct}{\mcitedefaultseppunct}\relax
\EndOfBibitem
\bibitem[Ramesh \latin{et~al.}(2022)Ramesh, Karuppasamy, Vikraman,
  Santhoshkumar, Bathula, Palem, Kathalingam, Kim, Kim, and Kim]{ramesh_2022}
Ramesh,~S.; Karuppasamy,~K.; Vikraman,~D.; Santhoshkumar,~P.; Bathula,~C.;
  Palem,~R.~R.; Kathalingam,~A.; Kim,~H.-S.; Kim,~J.-H.; Kim,~H.~S. Sheet-like
  morphology {CuCo2O4} bimetallic nanoparticles adorned on graphene oxide
  composites for symmetrical energy storage applications. \emph{Journal of
  Alloys and Compounds} \textbf{2022}, \emph{892}, 162182\relax
\mciteBstWouldAddEndPuncttrue
\mciteSetBstMidEndSepPunct{\mcitedefaultmidpunct}
{\mcitedefaultendpunct}{\mcitedefaultseppunct}\relax
\EndOfBibitem
\bibitem[Sabeeh \latin{et~al.}(2021)Sabeeh, Aadil, Zulfiqar, Rasheed,
  Al-Khalli, Agboola, Haider, Warsi, and Shakir]{sabeeh_2021}
Sabeeh,~H.; Aadil,~M.; Zulfiqar,~S.; Rasheed,~A.; Al-Khalli,~N.~F.;
  Agboola,~P.~O.; Haider,~S.; Warsi,~M.~F.; Shakir,~I. Hydrothermal synthesis
  of {CuS} nanochips and their nanohybrids with {CNTs} for electrochemical
  energy storage applications. \emph{Ceramics International} \textbf{2021},
  \emph{47}, 13613--13621\relax
\mciteBstWouldAddEndPuncttrue
\mciteSetBstMidEndSepPunct{\mcitedefaultmidpunct}
{\mcitedefaultendpunct}{\mcitedefaultseppunct}\relax
\EndOfBibitem
\bibitem[Sharma \latin{et~al.}(2019)Sharma, Kumar, Sharma, Naushad,
  Prakash~Dwivedi, ALOthman, and Mola]{sharma_2019}
Sharma,~G.; Kumar,~A.; Sharma,~S.; Naushad,~M.; Prakash~Dwivedi,~R.;
  ALOthman,~Z.~A.; Mola,~G.~T. Novel development of nanoparticles to bimetallic
  nanoparticles and their composites: {A} review. \emph{Journal of King Saud
  University - Science} \textbf{2019}, \emph{31}, 257--269\relax
\mciteBstWouldAddEndPuncttrue
\mciteSetBstMidEndSepPunct{\mcitedefaultmidpunct}
{\mcitedefaultendpunct}{\mcitedefaultseppunct}\relax
\EndOfBibitem
\bibitem[Idris and Roy(2023)Idris, and Roy]{idris_2023}
Idris,~D.~S.; Roy,~A. Synthesis of {Bimetallic} {Nanoparticles} and
  {Applications}—{An} {Updated} {Review}. \emph{Crystals} \textbf{2023},
  \emph{13}, 637\relax
\mciteBstWouldAddEndPuncttrue
\mciteSetBstMidEndSepPunct{\mcitedefaultmidpunct}
{\mcitedefaultendpunct}{\mcitedefaultseppunct}\relax
\EndOfBibitem
\bibitem[Coviello \latin{et~al.}()Coviello, Forrer, and
  Amendola]{coviello_2022}
Coviello,~V.; Forrer,~D.; Amendola,~V. Recent {{Developments}} in {{Plasmonic
  Alloy Nanoparticles}}: {{Synthesis}}, {{Modelling}}, {{Properties}} and
  {{Applications}}. \emph{23}, e202200136\relax
\mciteBstWouldAddEndPuncttrue
\mciteSetBstMidEndSepPunct{\mcitedefaultmidpunct}
{\mcitedefaultendpunct}{\mcitedefaultseppunct}\relax
\EndOfBibitem
\bibitem[Farkaš and family=Leeuw()Farkaš, and family=Leeuw]{farkas_2021}
Farkaš,~B.; family=Leeuw,~p.~u.,~given=Nora~H. A {{Perspective}} on
  {{Modelling Metallic Magnetic Nanoparticles}} in {{Biomedicine}}: {{From
  Monometals}} to {{Nanoalloys}} and {{Ligand-Protected Particles}}. \emph{14},
  3611\relax
\mciteBstWouldAddEndPuncttrue
\mciteSetBstMidEndSepPunct{\mcitedefaultmidpunct}
{\mcitedefaultendpunct}{\mcitedefaultseppunct}\relax
\EndOfBibitem
\bibitem[Ferrando()]{ferrando_2018}
Ferrando,~R. Determining the Equilibrium Structures of Nanoalloys by
  Computational Methods. \emph{20}, 179\relax
\mciteBstWouldAddEndPuncttrue
\mciteSetBstMidEndSepPunct{\mcitedefaultmidpunct}
{\mcitedefaultendpunct}{\mcitedefaultseppunct}\relax
\EndOfBibitem
\bibitem[family=Walle and Asta()family=Walle, and Asta]{walle_2019}
family=Walle,~p. d.~u.,~given=Axel; Asta,~M. High-Throughput Calculations in
  the Context of Alloy Design. \emph{44}, 252--256\relax
\mciteBstWouldAddEndPuncttrue
\mciteSetBstMidEndSepPunct{\mcitedefaultmidpunct}
{\mcitedefaultendpunct}{\mcitedefaultseppunct}\relax
\EndOfBibitem
\bibitem[Ferrando \latin{et~al.}()Ferrando, Jellinek, and
  Johnston]{ferrando_2008a}
Ferrando,~R.; Jellinek,~J.; Johnston,~R.~L. Nanoalloys:\, {{From Theory}} to
  {{Applications}} of {{Alloy Clusters}} and {{Nanoparticles}}. \emph{108},
  845--910\relax
\mciteBstWouldAddEndPuncttrue
\mciteSetBstMidEndSepPunct{\mcitedefaultmidpunct}
{\mcitedefaultendpunct}{\mcitedefaultseppunct}\relax
\EndOfBibitem
\bibitem[Ferrando()]{ferrando_2016}
Ferrando,~R. Theoretical and Computational Methods for Nanoalloy Structure and
  Thermodynamics. \emph{10}, 75--129\relax
\mciteBstWouldAddEndPuncttrue
\mciteSetBstMidEndSepPunct{\mcitedefaultmidpunct}
{\mcitedefaultendpunct}{\mcitedefaultseppunct}\relax
\EndOfBibitem
\bibitem[family=Rosa Abad \latin{et~al.}()family=Rosa Abad, Londoño-Calderon,
  Bringa, Soldano, family=Paz, Santiago, Mejía-Rosales, Yacamán, and
  Mariscal]{delarosaabad_2021}
family=Rosa Abad,~p. l.~u.,~given=Juan~A.; Londoño-Calderon,~A.;
  Bringa,~E.~M.; Soldano,~G.~J.; family=Paz,~g.-i.,~given=Sergio.~A.;
  Santiago,~U.; Mejía-Rosales,~S.~J.; Yacamán,~M.~J.; Mariscal,~M.~M. Soft or
  {{Hard}}? {{Investigating}} the {{Deformation Mechanisms}} of {{Au}}–{{Pd}}
  and {{Pd Nanocubes}} under {{Compression}}: {{An Experimental}} and
  {{Molecular Dynamics Study}}. \emph{125}, 25298--25306\relax
\mciteBstWouldAddEndPuncttrue
\mciteSetBstMidEndSepPunct{\mcitedefaultmidpunct}
{\mcitedefaultendpunct}{\mcitedefaultseppunct}\relax
\EndOfBibitem
\bibitem[Ferrando()]{ferrando_2015}
Ferrando,~R. Symmetry Breaking and Morphological Instabilities in Core-Shell
  Metallic Nanoparticles. \emph{27}, 13003\relax
\mciteBstWouldAddEndPuncttrue
\mciteSetBstMidEndSepPunct{\mcitedefaultmidpunct}
{\mcitedefaultendpunct}{\mcitedefaultseppunct}\relax
\EndOfBibitem
\bibitem[Palomares-Baez \latin{et~al.}()Palomares-Baez, Panizon, and
  Ferrando]{palomares-baez_2017}
Palomares-Baez,~J.~P.; Panizon,~E.; Ferrando,~R. Nanoscale {{Effects}} on
  {{Phase Separation}}. \emph{17}, 5394--5401\relax
\mciteBstWouldAddEndPuncttrue
\mciteSetBstMidEndSepPunct{\mcitedefaultmidpunct}
{\mcitedefaultendpunct}{\mcitedefaultseppunct}\relax
\EndOfBibitem
\bibitem[Nelli and Ferrando()Nelli, and Ferrando]{nelli_2019}
Nelli,~D.; Ferrando,~R. Core–Shell vs. Multi-Shell Formation in Nanoalloy
  Evolution from Disordered Configurations. \emph{11}, 13040--13050\relax
\mciteBstWouldAddEndPuncttrue
\mciteSetBstMidEndSepPunct{\mcitedefaultmidpunct}
{\mcitedefaultendpunct}{\mcitedefaultseppunct}\relax
\EndOfBibitem
\bibitem[Zhang \latin{et~al.}(2020)Zhang, Han, Zhu, Zhang, Li, Gao, Wu, Yang,
  Liu, Baaziz, Ersen, Gu, Miller, and Liu]{zhang_2020b}
Zhang,~X.; Han,~S.; Zhu,~B.; Zhang,~G.; Li,~X.; Gao,~Y.; Wu,~Z.; Yang,~B.;
  Liu,~Y.; Baaziz,~W.; Ersen,~O.; Gu,~M.; Miller,~J.~T.; Liu,~W. Reversible
  loss of core–shell structure for {Ni}–{Au} bimetallic nanoparticles
  during {CO2} hydrogenation. \emph{Nature Catalysis} \textbf{2020}, \emph{3},
  411--417, Number: 4 Publisher: Nature Publishing Group\relax
\mciteBstWouldAddEndPuncttrue
\mciteSetBstMidEndSepPunct{\mcitedefaultmidpunct}
{\mcitedefaultendpunct}{\mcitedefaultseppunct}\relax
\EndOfBibitem
\bibitem[Paz \latin{et~al.}()Paz, Leiva, Jellinek, and Mariscal]{paz_2011b}
Paz,~S.~A.; Leiva,~E. P.~M.; Jellinek,~J.; Mariscal,~M.~M. Properties of
  Rotating Nanoalloys Formed by Cluster Collision: {{A}} Computer Simulation
  Study. \emph{134}, 094701\relax
\mciteBstWouldAddEndPuncttrue
\mciteSetBstMidEndSepPunct{\mcitedefaultmidpunct}
{\mcitedefaultendpunct}{\mcitedefaultseppunct}\relax
\EndOfBibitem
\bibitem[Paz and Leiva()Paz, and Leiva]{paz_2014}
Paz,~S.~A.; Leiva,~E. P.~M. Unveiling the Mechanism of Core–Shell Formation
  by Counting the Relative Occurrence of Microstates. \emph{595--596},
  87--90\relax
\mciteBstWouldAddEndPuncttrue
\mciteSetBstMidEndSepPunct{\mcitedefaultmidpunct}
{\mcitedefaultendpunct}{\mcitedefaultseppunct}\relax
\EndOfBibitem
\bibitem[Farigliano \latin{et~al.}()Farigliano, Villarreal, Leiva, and
  Paz]{farigliano_2020}
Farigliano,~L.~M.; Villarreal,~M.~A.; Leiva,~E.~P.; Paz,~S.~A. Thermodynamics
  of {{Nanoparticle Coalescence}} at {{Different Temperatures}} via
  {{Well-Tempered Metadynamics}}. \emph{124}, 24009--24016\relax
\mciteBstWouldAddEndPuncttrue
\mciteSetBstMidEndSepPunct{\mcitedefaultmidpunct}
{\mcitedefaultendpunct}{\mcitedefaultseppunct}\relax
\EndOfBibitem
\bibitem[Farigliano \latin{et~al.}()Farigliano, Paz, Leiva, and
  Villarreal]{farigliano_2017}
Farigliano,~L.~M.; Paz,~S.~A.; Leiva,~E. P.~M.; Villarreal,~M. M.~A.
  Coalescence of {{Nanoclusters Analyzed}} by {{Well-Tempered Metadynamics}}.
  {{Comparison}} with {{Straightforward Molecular Dynamics}}. \emph{13},
  3874--3880\relax
\mciteBstWouldAddEndPuncttrue
\mciteSetBstMidEndSepPunct{\mcitedefaultmidpunct}
{\mcitedefaultendpunct}{\mcitedefaultseppunct}\relax
\EndOfBibitem
\bibitem[Goudeli()]{goudeli_2019}
Goudeli,~E. Nanoparticle Growth, Coalescence, and Phase Change in the Gas-Phase
  by Molecular Dynamics. \emph{23}, 155--163\relax
\mciteBstWouldAddEndPuncttrue
\mciteSetBstMidEndSepPunct{\mcitedefaultmidpunct}
{\mcitedefaultendpunct}{\mcitedefaultseppunct}\relax
\EndOfBibitem
\bibitem[Grammatikopoulos \latin{et~al.}()Grammatikopoulos, Sowwan, and
  Kioseoglou]{grammatikopoulos_2019}
Grammatikopoulos,~P.; Sowwan,~M.; Kioseoglou,~J. Computational {{Modeling}} of
  {{Nanoparticle Coalescence}}. 1900013\relax
\mciteBstWouldAddEndPuncttrue
\mciteSetBstMidEndSepPunct{\mcitedefaultmidpunct}
{\mcitedefaultendpunct}{\mcitedefaultseppunct}\relax
\EndOfBibitem
\bibitem[Paz and Leiva()Paz, and Leiva]{paz_2015a}
Paz,~S.~A.; Leiva,~E. P.~M. Time {{Recovery}} for a {{Complex Process Using
  Accelerated Dynamics}}. \emph{11}, 1725--1734\relax
\mciteBstWouldAddEndPuncttrue
\mciteSetBstMidEndSepPunct{\mcitedefaultmidpunct}
{\mcitedefaultendpunct}{\mcitedefaultseppunct}\relax
\EndOfBibitem
\bibitem[Elstner \latin{et~al.}()Elstner, Porezag, Jungnickel, Elsner, Haugk,
  family=Frauenheim, Suhai, and Seifert]{elstner_1998}
Elstner,~M.; Porezag,~D.; Jungnickel,~G.; Elsner,~J.; Haugk,~M.;
  family=Frauenheim,~g.-i.,~given=Th.; Suhai,~S.; Seifert,~G.
  Self-Consistent-Charge Density-Functional Tight-Binding Method for
  Simulations of Complex Materials Properties. \emph{58}, 7260--7268\relax
\mciteBstWouldAddEndPuncttrue
\mciteSetBstMidEndSepPunct{\mcitedefaultmidpunct}
{\mcitedefaultendpunct}{\mcitedefaultseppunct}\relax
\EndOfBibitem
\bibitem[family=Frauenheim \latin{et~al.}()family=Frauenheim, Seifert,
  Elsterner, Hajnal, Jungnickel, Porezag, Suhai, and Scholz]{frauenheim_2000}
family=Frauenheim,~g.-i.,~given=Th.; Seifert,~G.; Elsterner,~M.; Hajnal,~Z.;
  Jungnickel,~G.; Porezag,~D.; Suhai,~S.; Scholz,~R. A {{Self-Consistent Charge
  Density-Functional Based Tight-Binding Method}} for {{Predictive Materials
  Simulations}} in {{Physics}}, {{Chemistry}} and {{Biology}}. \emph{217},
  41--62\relax
\mciteBstWouldAddEndPuncttrue
\mciteSetBstMidEndSepPunct{\mcitedefaultmidpunct}
{\mcitedefaultendpunct}{\mcitedefaultseppunct}\relax
\EndOfBibitem
\bibitem[Seifert()]{seifert_2007}
Seifert,~G. Tight-{{Binding Density Functional Theory}}:\, {{An Approximate
  Kohn}} {{Sham DFT Scheme}}. \emph{111}, 5609--5613\relax
\mciteBstWouldAddEndPuncttrue
\mciteSetBstMidEndSepPunct{\mcitedefaultmidpunct}
{\mcitedefaultendpunct}{\mcitedefaultseppunct}\relax
\EndOfBibitem
\bibitem[Gaus \latin{et~al.}()Gaus, Cui, and Elstner]{gaus_2011}
Gaus,~M.; Cui,~Q.; Elstner,~M. {{DFTB3}}: {{Extension}} of the
  {{Self-Consistent-Charge Density-Functional Tight-Binding Method}}
  ({{SCC-DFTB}}). \emph{7}, 931--948\relax
\mciteBstWouldAddEndPuncttrue
\mciteSetBstMidEndSepPunct{\mcitedefaultmidpunct}
{\mcitedefaultendpunct}{\mcitedefaultseppunct}\relax
\EndOfBibitem
\bibitem[Van~den Bossche()]{vandenbossche_2019a}
Van~den Bossche,~M. Hotcent. \url{https://gitlab.com/mvdb/hotcent}\relax
\mciteBstWouldAddEndPuncttrue
\mciteSetBstMidEndSepPunct{\mcitedefaultmidpunct}
{\mcitedefaultendpunct}{\mcitedefaultseppunct}\relax
\EndOfBibitem
\bibitem[Van~den Bossche()]{vandenbossche_2019}
Van~den Bossche,~M. {{DFTB-Assisted Global Structure Optimization}} of 13- and
  55-{{Atom Late Transition Metal Clusters}}. \emph{123}, 3038--3045\relax
\mciteBstWouldAddEndPuncttrue
\mciteSetBstMidEndSepPunct{\mcitedefaultmidpunct}
{\mcitedefaultendpunct}{\mcitedefaultseppunct}\relax
\EndOfBibitem
\bibitem[Larsen \latin{et~al.}()Larsen, Mortensen, Blomqvist, Castelli,
  Christensen, Dułak, Friis, Groves, Hammer, Hargus, Hermes, Jennings, Jensen,
  Kermode, Kitchin, Kolsbjerg, Kubal, Kaasbjerg, Lysgaard, Maronsson, Maxson,
  Olsen, Pastewka, Peterson, Rostgaard, Schiøtz, Schütt, Strange, Thygesen,
  Vegge, Vilhelmsen, Walter, Zeng, and Jacobsen]{larsen_2017}
Larsen,~A.~H. \latin{et~al.}  The Atomic Simulation Environment—a {{Python}}
  Library for Working with Atoms. \emph{29}, 273002\relax
\mciteBstWouldAddEndPuncttrue
\mciteSetBstMidEndSepPunct{\mcitedefaultmidpunct}
{\mcitedefaultendpunct}{\mcitedefaultseppunct}\relax
\EndOfBibitem
\bibitem[Enkovaara \latin{et~al.}()Enkovaara, Rostgaard, Mortensen, Chen,
  Dułak, Ferrighi, Gavnholt, Glinsvad, Haikola, Hansen, Kristoffersen, Kuisma,
  Larsen, Lehtovaara, Ljungberg, Lopez-Acevedo, Moses, Ojanen, Olsen, Petzold,
  Romero, Stausholm-Møller, Strange, Tritsaris, Vanin, Walter, Hammer,
  Häkkinen, Madsen, Nieminen, Nørskov, Puska, Rantala, Schiøtz, Thygesen,
  and Jacobsen]{enkovaara_2010}
Enkovaara,~J. \latin{et~al.}  Electronic Structure Calculations with {{GPAW}}:
  A Real-Space Implementation of the Projector Augmented-Wave Method.
  \emph{22}, 253202\relax
\mciteBstWouldAddEndPuncttrue
\mciteSetBstMidEndSepPunct{\mcitedefaultmidpunct}
{\mcitedefaultendpunct}{\mcitedefaultseppunct}\relax
\EndOfBibitem
\bibitem[Mortensen \latin{et~al.}()Mortensen, Hansen, and
  Jacobsen]{mortensen_2005}
Mortensen,~J.~J.; Hansen,~L.~B.; Jacobsen,~K.~W. Real-Space Grid Implementation
  of the Projector Augmented Wave Method. \emph{71}, 035109\relax
\mciteBstWouldAddEndPuncttrue
\mciteSetBstMidEndSepPunct{\mcitedefaultmidpunct}
{\mcitedefaultendpunct}{\mcitedefaultseppunct}\relax
\EndOfBibitem
\bibitem[Blöchl()]{blochl_1994}
Blöchl,~P.~E. Projector Augmented-Wave Method. \emph{50}, 17953--17979\relax
\mciteBstWouldAddEndPuncttrue
\mciteSetBstMidEndSepPunct{\mcitedefaultmidpunct}
{\mcitedefaultendpunct}{\mcitedefaultseppunct}\relax
\EndOfBibitem
\bibitem[Jain \latin{et~al.}()Jain, Ong, Hautier, Chen, Richards, Dacek,
  Cholia, Gunter, Skinner, Ceder, and Persson]{jain_2013}
Jain,~A.; Ong,~S.~P.; Hautier,~G.; Chen,~W.; Richards,~W.~D.; Dacek,~S.;
  Cholia,~S.; Gunter,~D.; Skinner,~D.; Ceder,~G.; Persson,~K.~A. Commentary:
  {{The Materials Project}}: {{A}} Materials Genome Approach to Accelerating
  Materials Innovation. \emph{1}, 011002\relax
\mciteBstWouldAddEndPuncttrue
\mciteSetBstMidEndSepPunct{\mcitedefaultmidpunct}
{\mcitedefaultendpunct}{\mcitedefaultseppunct}\relax
\EndOfBibitem
\bibitem[Fernandes \latin{et~al.}(2018)Fernandes, Machado, and
  Ferr{\~a}o]{fernandes2018quantitative}
Fernandes,~G.~F.; Machado,~F.~B.; Ferr{\~a}o,~L.~F. A quantitative tool to
  establish magic number clusters, $\varepsilon$3, applied in small silicon
  clusters, Si2-11. \emph{Journal of Molecular Modeling} \textbf{2018},
  \emph{24}, 1--9\relax
\mciteBstWouldAddEndPuncttrue
\mciteSetBstMidEndSepPunct{\mcitedefaultmidpunct}
{\mcitedefaultendpunct}{\mcitedefaultseppunct}\relax
\EndOfBibitem
\bibitem[Wang \latin{et~al.}(2005)Wang, Zhou, Wang, and
  Zhao]{wang2005optimally}
Wang,~J.; Zhou,~X.; Wang,~G.; Zhao,~J. Optimally stuffed fullerene structures
  of silicon nanoclusters. \emph{Physical Review B} \textbf{2005}, \emph{71},
  113412\relax
\mciteBstWouldAddEndPuncttrue
\mciteSetBstMidEndSepPunct{\mcitedefaultmidpunct}
{\mcitedefaultendpunct}{\mcitedefaultseppunct}\relax
\EndOfBibitem
\bibitem[Jain \latin{et~al.}(2013)Jain, Ong, Hautier, Chen, Richards, Dacek,
  Cholia, Gunter, Skinner, Ceder, and Persson]{Jain2013}
Jain,~A.; Ong,~S.~P.; Hautier,~G.; Chen,~W.; Richards,~W.~D.; Dacek,~S.;
  Cholia,~S.; Gunter,~D.; Skinner,~D.; Ceder,~G.; Persson,~K.~a. {The Materials
  Project: A materials genome approach to accelerating materials innovation}.
  \emph{APL Materials} \textbf{2013}, \emph{1}, 011002\relax
\mciteBstWouldAddEndPuncttrue
\mciteSetBstMidEndSepPunct{\mcitedefaultmidpunct}
{\mcitedefaultendpunct}{\mcitedefaultseppunct}\relax
\EndOfBibitem
\bibitem[Paz()]{paz_2020b}
Paz,~S.~A. {{GEMS}} Is an {{Extensible Molecular Simulator}}.
  \url{https://github.com/alexispaz/GEMS}\relax
\mciteBstWouldAddEndPuncttrue
\mciteSetBstMidEndSepPunct{\mcitedefaultmidpunct}
{\mcitedefaultendpunct}{\mcitedefaultseppunct}\relax
\EndOfBibitem
\bibitem[Hourahine \latin{et~al.}()Hourahine, Aradi, Blum, Bonafé, Buccheri,
  Camacho, Cevallos, Deshaye, Dumitrică, Dominguez, Ehlert, Elstner,
  family=Heide, Hermann, Irle, Kranz, Köhler, Kowalczyk, Kubař, Lee, Lutsker,
  Maurer, Min, Mitchell, Negre, Niehaus, Niklasson, Page, Pecchia, Penazzi,
  Persson, Řezáč, Sánchez, Sternberg, Stöhr, Stuckenberg, Tkatchenko, Yu,
  and Frauenheim]{hourahine_2020}
Hourahine,~B. \latin{et~al.}  {{DFTB}}+, a Software Package for Efficient
  Approximate Density Functional Theory Based Atomistic Simulations.
  \emph{152}, 124101\relax
\mciteBstWouldAddEndPuncttrue
\mciteSetBstMidEndSepPunct{\mcitedefaultmidpunct}
{\mcitedefaultendpunct}{\mcitedefaultseppunct}\relax
\EndOfBibitem
\bibitem[Chattopadhyay \latin{et~al.}(2001)Chattopadhyay, Manna, Talapatra, and
  Pabi]{chattopadhyay2001mathematical}
Chattopadhyay,~P.; Manna,~I.; Talapatra,~S.; Pabi,~S. A mathematical analysis
  of milling mechanics in a planetary ball mill. \emph{Materials Chemistry and
  Physics} \textbf{2001}, \emph{68}, 85--94\relax
\mciteBstWouldAddEndPuncttrue
\mciteSetBstMidEndSepPunct{\mcitedefaultmidpunct}
{\mcitedefaultendpunct}{\mcitedefaultseppunct}\relax
\EndOfBibitem
\bibitem[Humphrey \latin{et~al.}(1996)Humphrey, Dalke, and Schulten]{HUMP96}
Humphrey,~W.; Dalke,~A.; Schulten,~K. {VMD} -- {V}isual {M}olecular {D}ynamics.
  \emph{Journal of Molecular Graphics} \textbf{1996}, \emph{14}, 33--38\relax
\mciteBstWouldAddEndPuncttrue
\mciteSetBstMidEndSepPunct{\mcitedefaultmidpunct}
{\mcitedefaultendpunct}{\mcitedefaultseppunct}\relax
\EndOfBibitem
\bibitem[Varshney \latin{et~al.}(1994)Varshney, Brooks, and Wright]{VARSH1994}
Varshney,~A.; Brooks,~F.~P.; Wright,~W.~V. Linearly Scalable Computation of
  Smooth Molecular Surfaces. \emph{IEEE Computer Graphics and Applications}
  \textbf{1994}, \emph{14}, 19--25\relax
\mciteBstWouldAddEndPuncttrue
\mciteSetBstMidEndSepPunct{\mcitedefaultmidpunct}
{\mcitedefaultendpunct}{\mcitedefaultseppunct}\relax
\EndOfBibitem
\bibitem[Oviedo \latin{et~al.}()Oviedo, Fernandez, Otero, Leiva, and
  Paz]{oviedo_2023}
Oviedo,~M.~B.; Fernandez,~F.; Otero,~M.; Leiva,~E. P.~M.; Paz,~S.~A. Density
  {{Functional Tight-Binding Model}} for {{Lithium}}–{{Silicon Alloys}}.
  \relax
\mciteBstWouldAddEndPunctfalse
\mciteSetBstMidEndSepPunct{\mcitedefaultmidpunct}
{}{\mcitedefaultseppunct}\relax
\EndOfBibitem
\bibitem[Van~den Bossche \latin{et~al.}()Van~den Bossche, Grönbeck, and
  Hammer]{vandenbossche_2018}
Van~den Bossche,~M.; Grönbeck,~H.; Hammer,~B. Tight-{{Binding
  Approximation-Enhanced Global Optimization}}. \emph{14}, 2797--2807\relax
\mciteBstWouldAddEndPuncttrue
\mciteSetBstMidEndSepPunct{\mcitedefaultmidpunct}
{\mcitedefaultendpunct}{\mcitedefaultseppunct}\relax
\EndOfBibitem
\bibitem[Zhang \latin{et~al.}()Zhang, Ying, Quan, Shi, and Li]{zhang_2012a}
Zhang,~X.; Ying,~C.; Quan,~S.; Shi,~G.; Li,~Z. A First Principles Investigation
  on the Structural, Phonon, Elastic and Thermodynamic Properties of the
  {{Si0}}.{{5Sn0}}.5 Cubic Alloy. \emph{152}, 955--959\relax
\mciteBstWouldAddEndPuncttrue
\mciteSetBstMidEndSepPunct{\mcitedefaultmidpunct}
{\mcitedefaultendpunct}{\mcitedefaultseppunct}\relax
\EndOfBibitem
\bibitem[Eom \latin{et~al.}(2021)Eom, Messing, Johansson, and
  Deppert]{eom2021general}
Eom,~N.; Messing,~M.~E.; Johansson,~J.; Deppert,~K. General trends in
  core--shell preferences for bimetallic nanoparticles. \emph{ACS nano}
  \textbf{2021}, \emph{15}, 8883--8895\relax
\mciteBstWouldAddEndPuncttrue
\mciteSetBstMidEndSepPunct{\mcitedefaultmidpunct}
{\mcitedefaultendpunct}{\mcitedefaultseppunct}\relax
\EndOfBibitem
\bibitem[Wang and Johnson(2009)Wang, and Johnson]{wang2009predicted}
Wang,~L.-L.; Johnson,~D.~D. Predicted trends of core- shell preferences for 132
  late transition-metal binary-alloy nanoparticles. \emph{Journal of the
  American Chemical Society} \textbf{2009}, \emph{131}, 14023--14029\relax
\mciteBstWouldAddEndPuncttrue
\mciteSetBstMidEndSepPunct{\mcitedefaultmidpunct}
{\mcitedefaultendpunct}{\mcitedefaultseppunct}\relax
\EndOfBibitem
\bibitem[Kittel(2005)]{kittel2005introduction}
Kittel,~C. Introduction to solid state physics, John Wiley \& Sons. \emph{Inc.,
  Sixth edition,(New York, 1986)} \textbf{2005}, \relax
\mciteBstWouldAddEndPunctfalse
\mciteSetBstMidEndSepPunct{\mcitedefaultmidpunct}
{}{\mcitedefaultseppunct}\relax
\EndOfBibitem
\end{mcitethebibliography}

\begin{widetext}
    \begin{center}
        \textbf{\large{SUPPLEMENTAL MATERIAL}}
    \end{center}

\begin{figure}[h!]
\includegraphics[width=6cm]{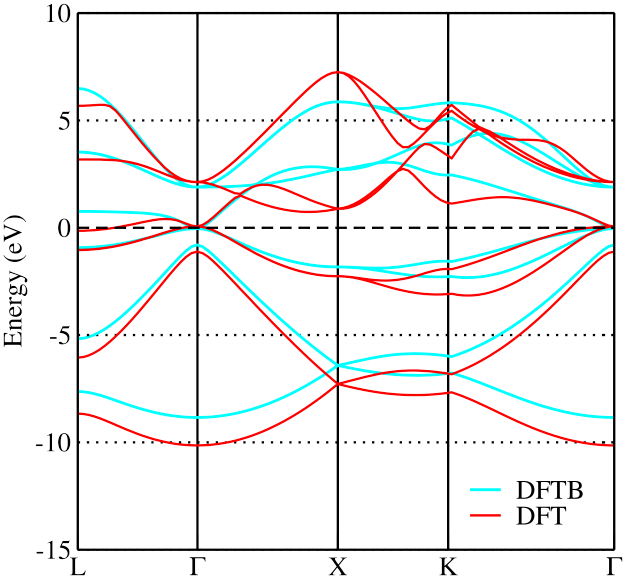}\hspace{.5cm}
\caption{Band structure of Sn computed by DFTB, in comparison with the band structures computed by DFT/PBE. The electronic bands are shifted to the respective Fermi levels (0 eV).
}
\label{bands}
\end{figure}

\begin{figure}[tbh!]
\center
\subfigure[SASA, $n_{\text{Si}}$=5.]{\includegraphics[width=0.4\linewidth]{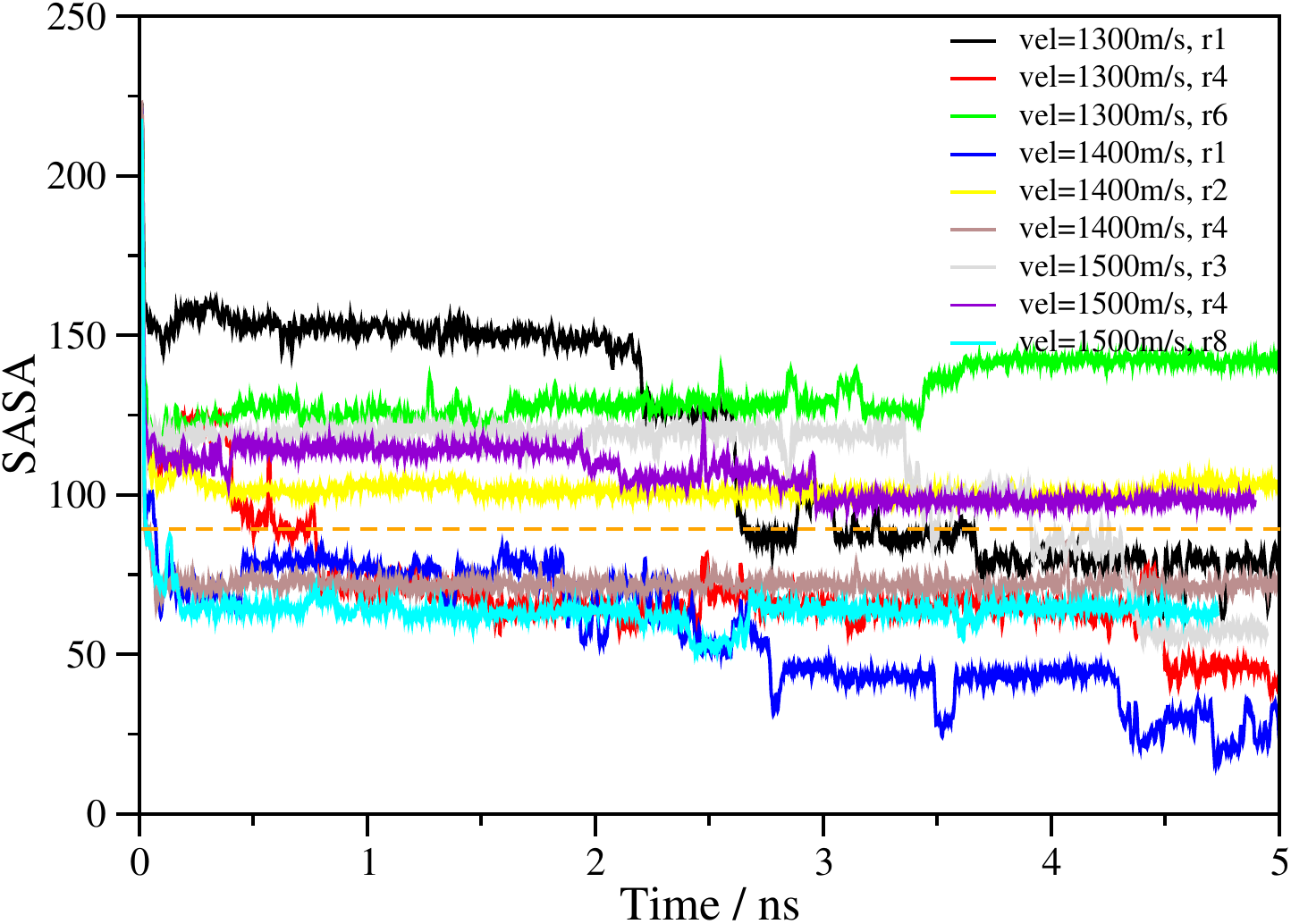}}
\subfigure[SASA, $n_{\text{Si}}$=10.]{\includegraphics[width=0.4\linewidth]{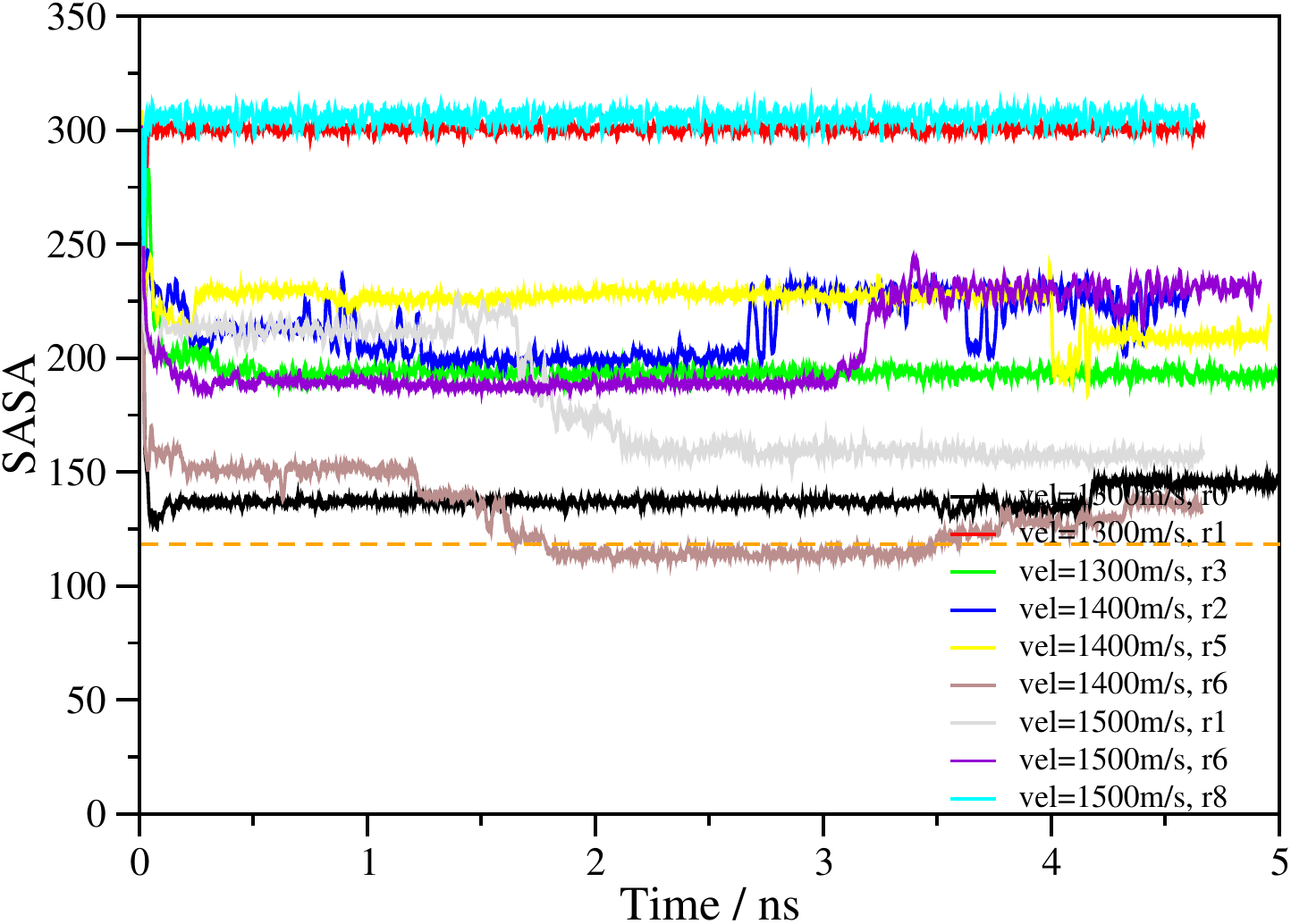}}
\subfigure[SASA, $n_{\text{Si}}$=11.]{\includegraphics[width=0.4\linewidth]{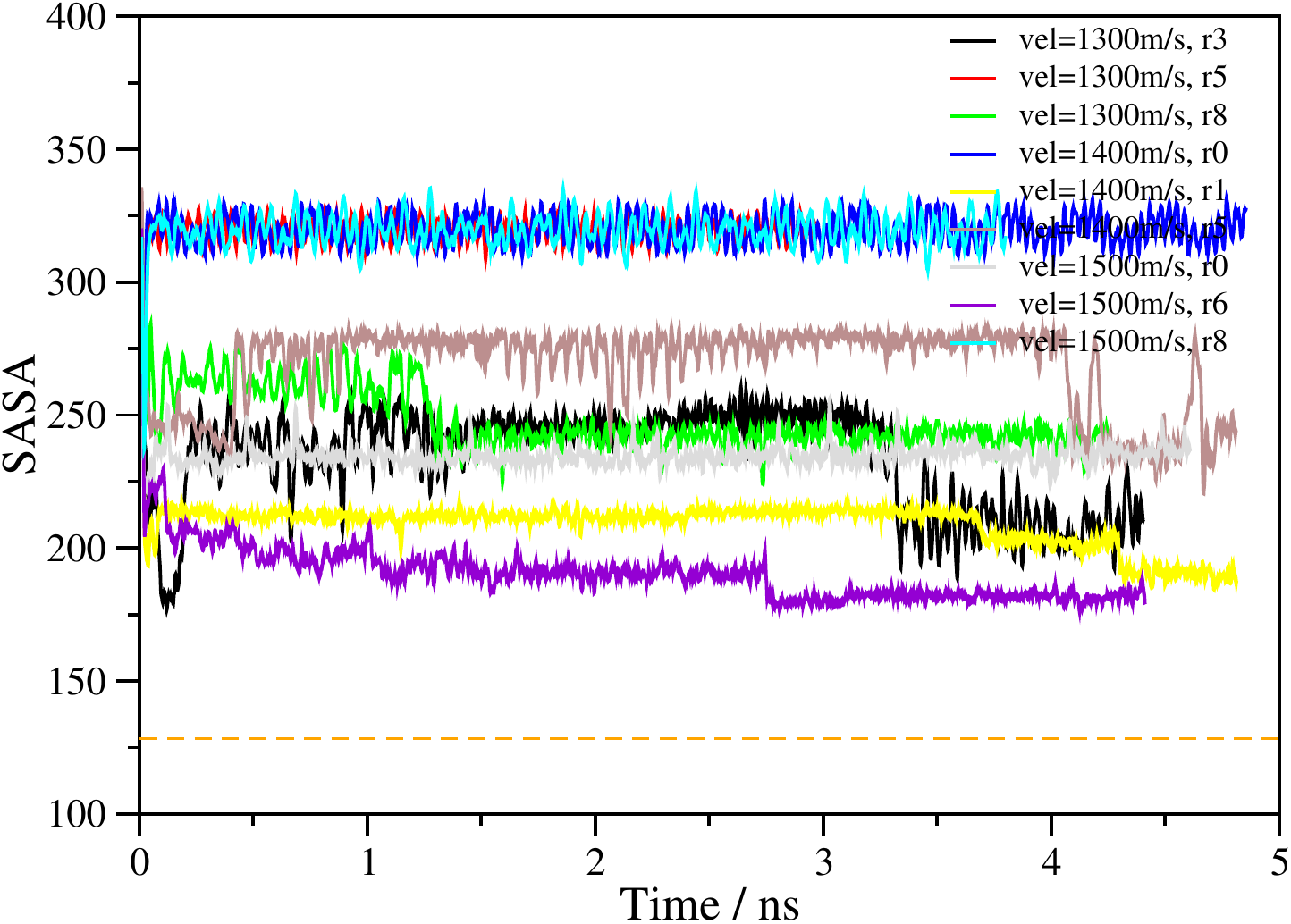}}
\caption{All SASA of collisions of Si clusters with different numbers of atoms against a 110 Sn cluster.}
\label{reg}
\end{figure}

\begin{figure}[tbh!]
\center
\subfigure[Si RMSD, $n_{\text{Si}}$=5.]{\includegraphics[width=0.8\linewidth]{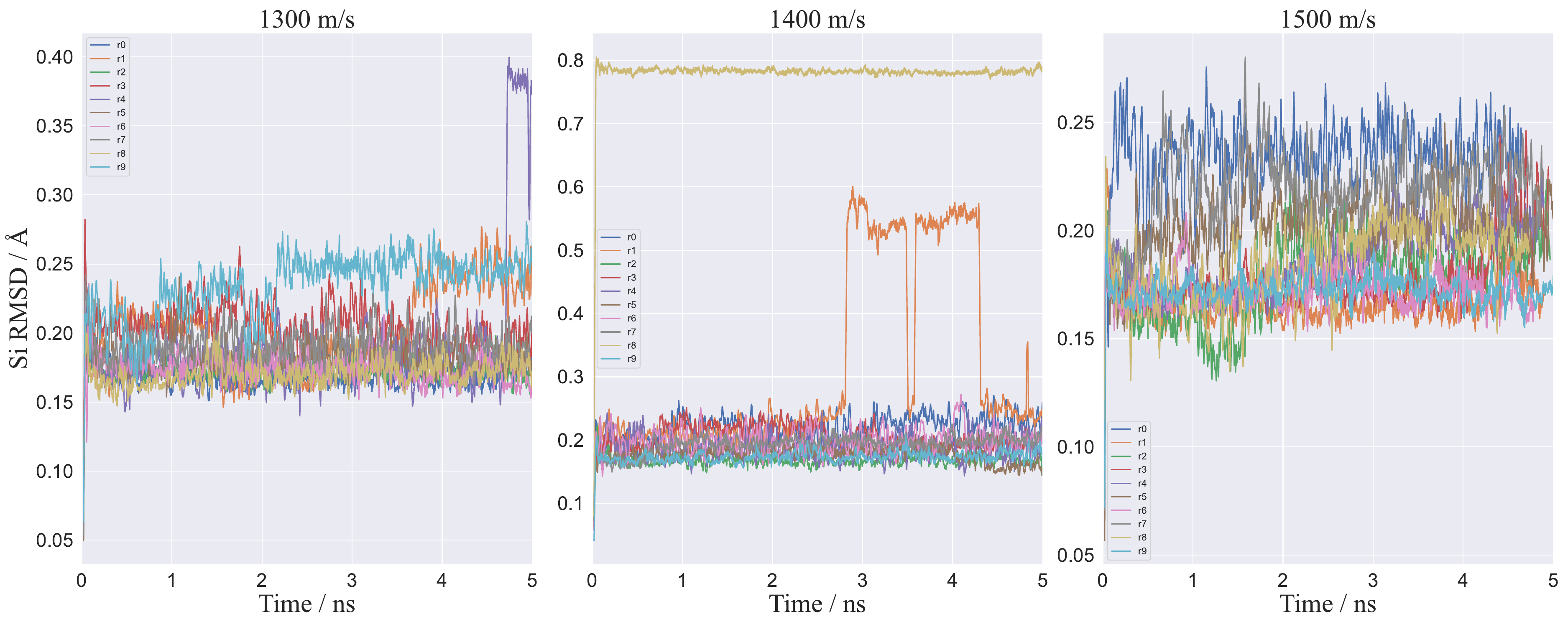}}
\subfigure[Si RMSD, $n_{\text{Si}}$=10.]{\includegraphics[width=0.8\linewidth]{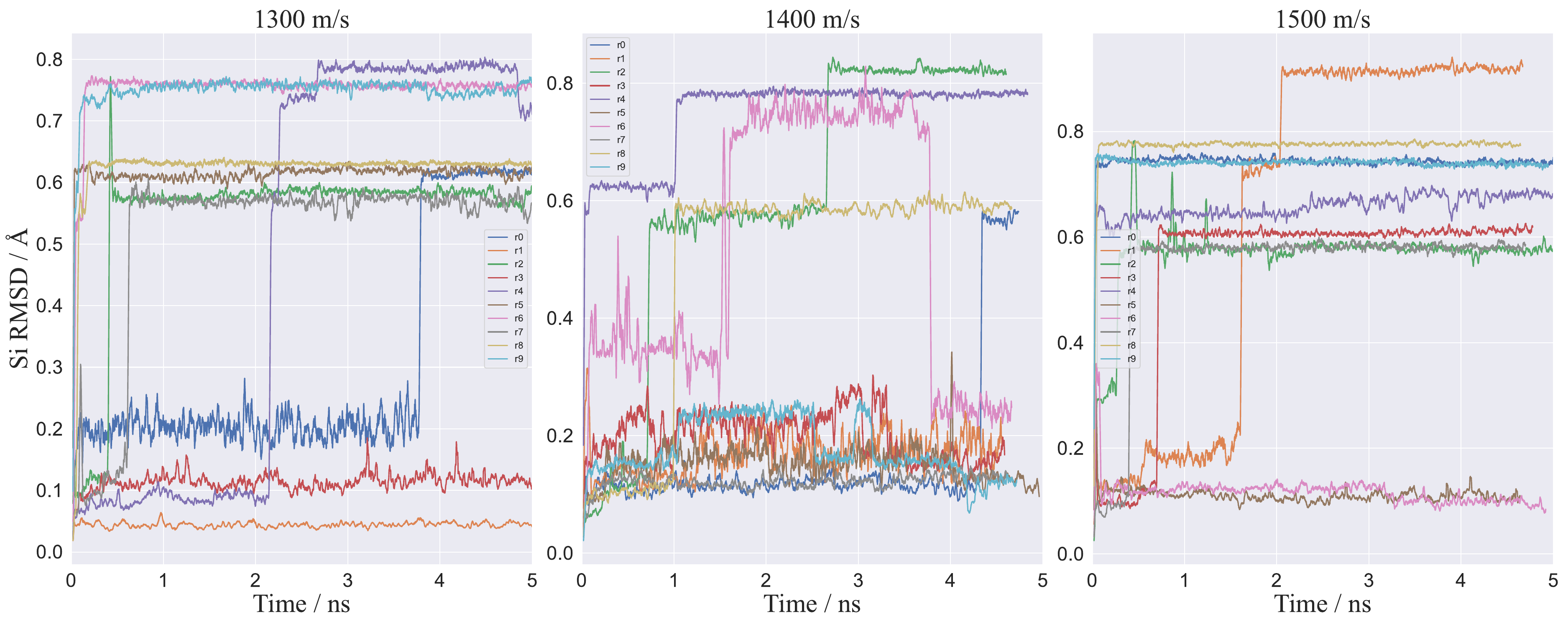}}
\subfigure[Si RMSD, $n_{\text{Si}}$=11.]{\includegraphics[width=0.9\linewidth]{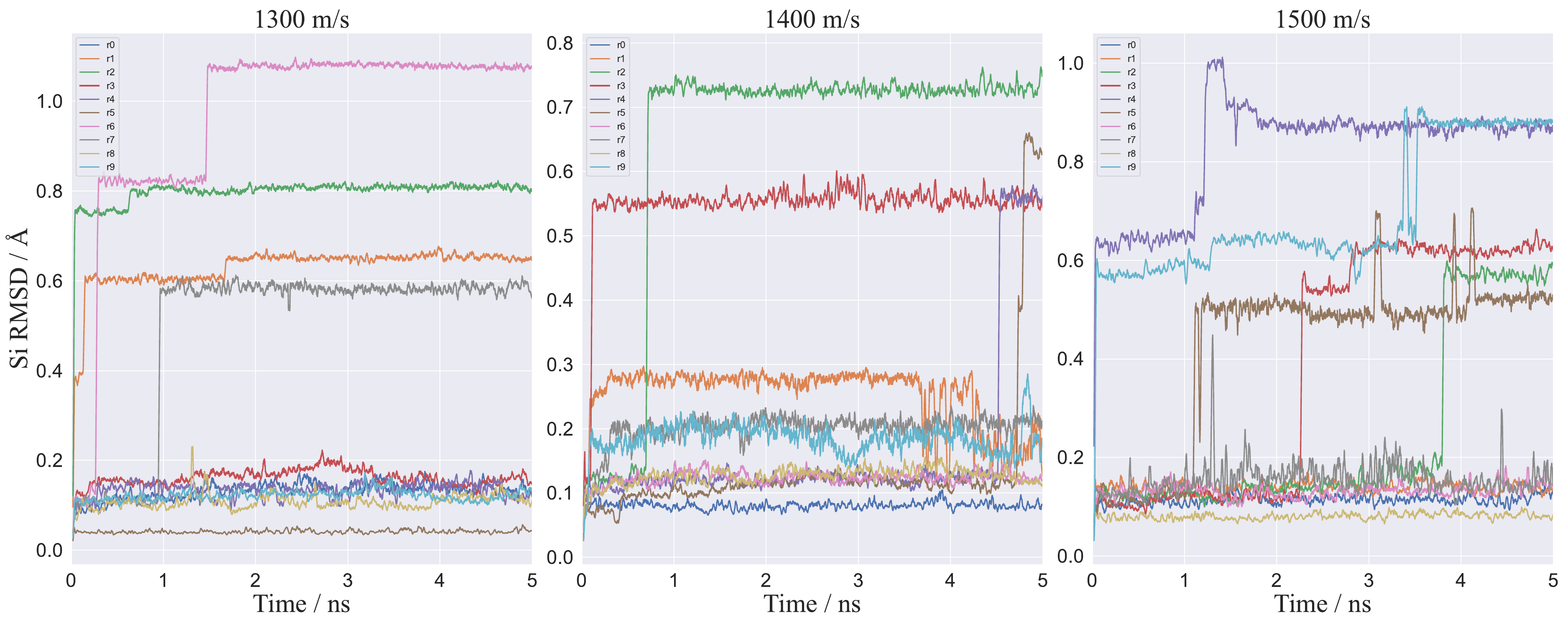}}
\caption{All Si RMSD of collisions of Si clusters with different numbers of atoms against a 110 Sn cluster.}
\label{reg}
\end{figure}

\begin{figure}[tbh!]
\center
\subfigure[Relative kinetic energy, $n_{\text{Si}}$=5.]{\includegraphics[width=0.8\linewidth]{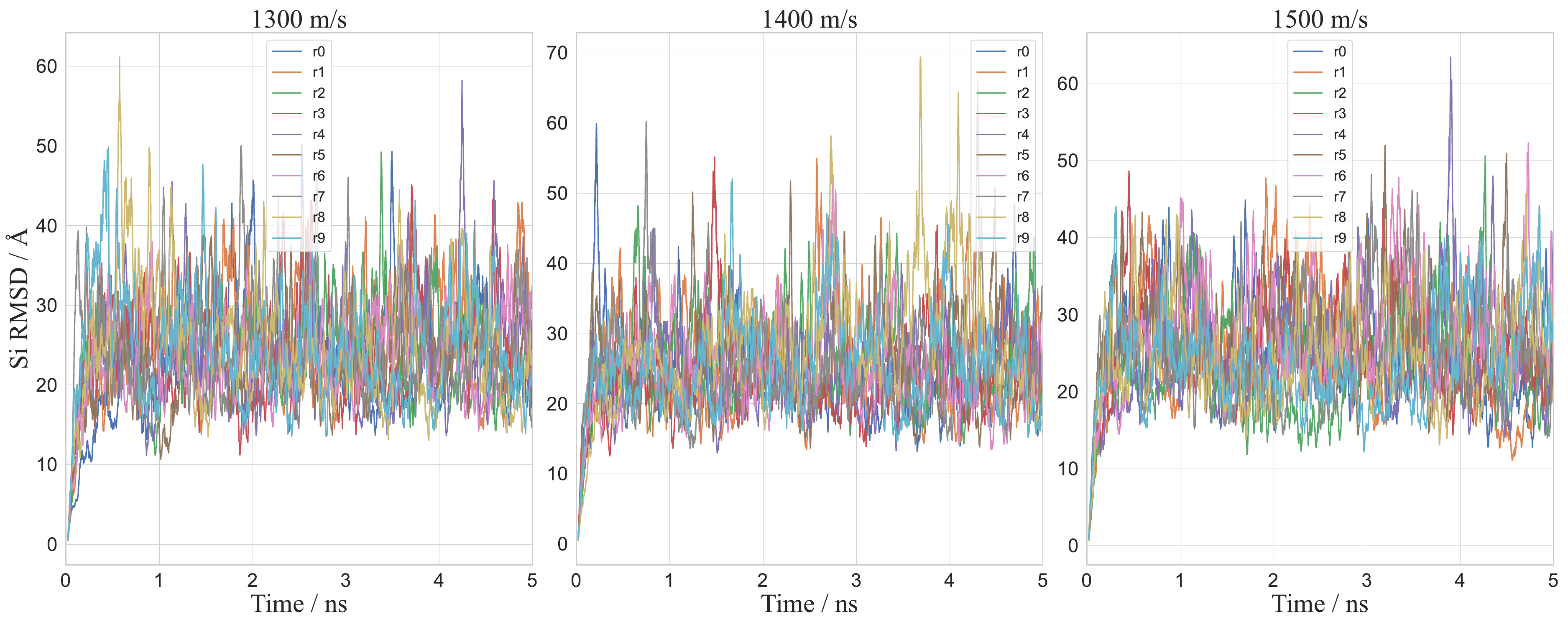}}
\subfigure[Relative kinetic energy, $n_{\text{Si}}$=10.]{\includegraphics[width=0.8\linewidth]{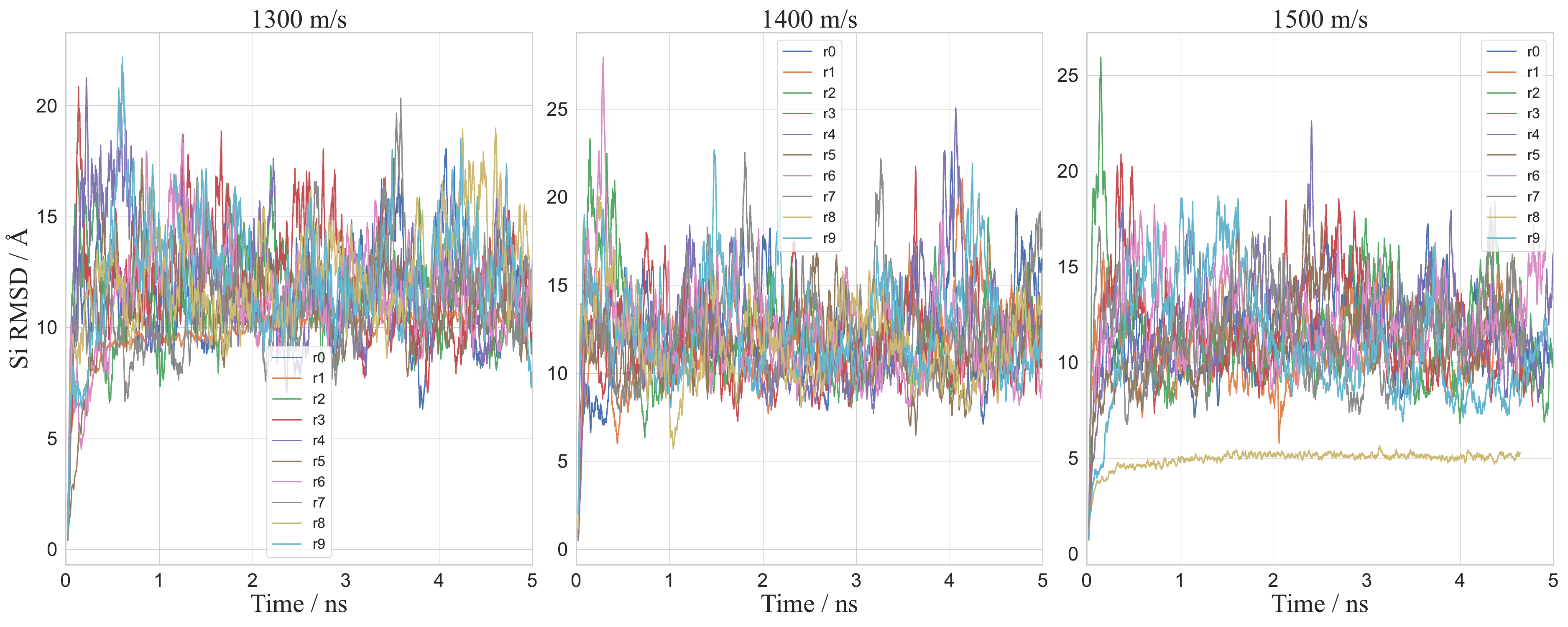}}
\subfigure[Relative kinetic energy, $n_{\text{Si}}$=11.]{\includegraphics[width=0.9\linewidth]{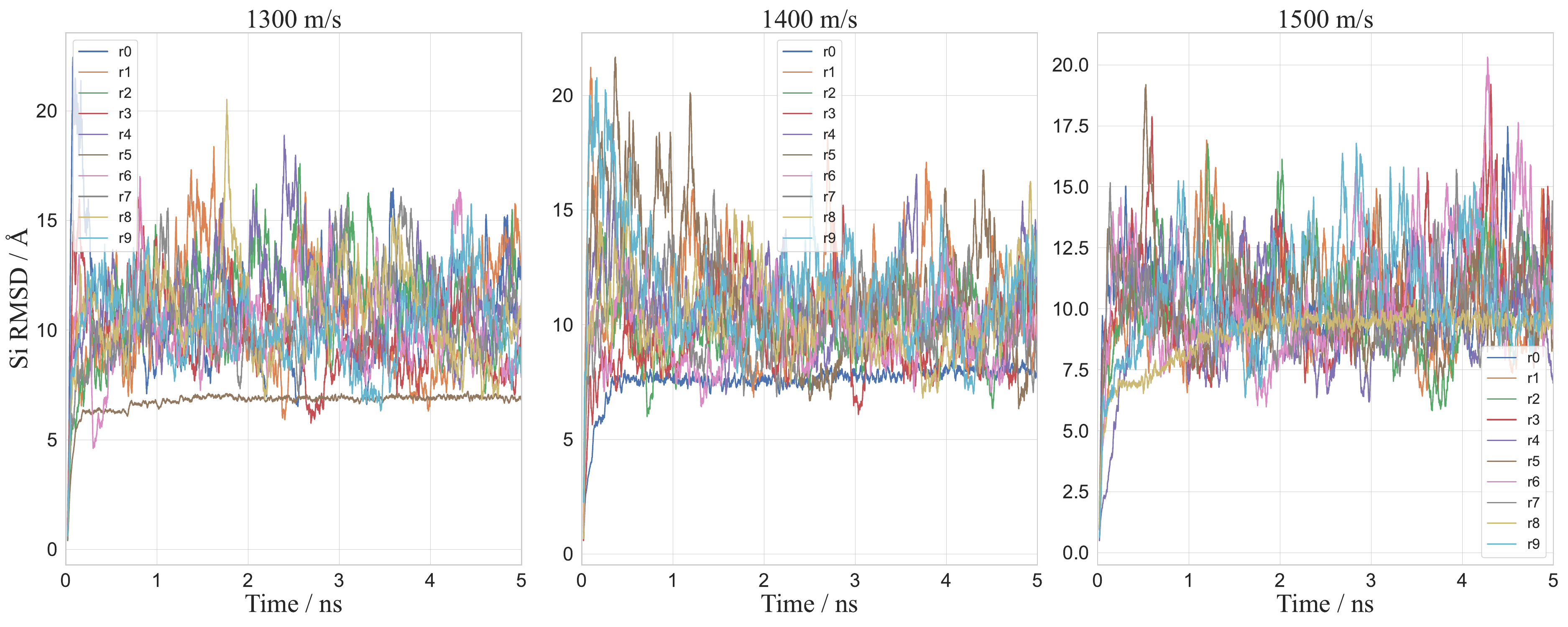}}
\caption{Relative kinetic energy between Sn and Si clusters}
\label{reg}
\end{figure}

\end{widetext}

\end{document}